\documentclass[3p]{elsarticle}

\journal{International Journal of Rock Mechanics and Mining Sciences}

\usepackage{framed,multirow}
\usepackage{amssymb}
\usepackage{latexsym}
\usepackage{amsmath}
\usepackage{rotating}
\usepackage{graphicx,color}
\usepackage{bm}
\usepackage{color}
\usepackage{multirow}
\usepackage{algorithm}
\usepackage{algpseudocode}
\usepackage{enumitem}
\usepackage{breakcites}		
\usepackage{natbib}
\usepackage{lscape}
\usepackage{array}
\usepackage{longtable}
\usepackage{epstopdf}
\usepackage{morefloats}
\usepackage{algpseudocode}
\usepackage{subcaption}
\usepackage{url}
\usepackage{xcolor}
\usepackage{booktabs}
\usepackage{soul}
\usepackage{lineno}
\usepackage{blindtext}

\begin{document}

\begin{frontmatter}
	\title{A multi-step calibration strategy for reliable parameter determination of salt rock mechanics constitutive models}%
	
	\author[TUDelftCiTG]{Hermínio T. Honório\corref{cor1}\fnref{fn1}}
	\author[Shell]{Maartje Houben}
	\author[Shell]{Kevin Bisdom}
	\author[Shell]{Arjan van der Linden}
	\author[Shell]{Karin de Borst}
	\author[TUDelftCiTG]{Lambertus J. Sluys}	
	\author[TUDelftCiTG]{Hadi Hajibeygi}
        
	\cortext[cor1]{Corresponding author}
	\fntext[fn1]{H.TasinafoHonorio@tudelft.nl}

	\address[TUDelftCiTG]{Faculty of Civil Engineering and Geosciences, Delft University of Technology, Stevinweg 1, 2628CN, Delft, The Netherlands}
	\address[Shell]{Shell Global Solutions International B.V., Grasweg 31, 1031 HW Amsterdam, The Netherlands}
 
	\begin{abstract}
            The storage of renewable hydrogen in salt caverns requires fast injection and production rates to cope with the imbalance between energy production and consumption. Such operational conditions raise concerns about the mechanical stability of salt caverns. Choosing an appropriate constitutive model for salt mechanics is an important step in investigating this issue, and many constitutive models with several parameters have been presented in the literature. However, a robust calibration strategy to reliably determine which model and which parameter set represent the given rock, based on stress-strain data, remains an unsolved challenge. For the first time in the community, we present a multi-step strategy to determine a single parameter set based on many deformation datasets for salt rocks. Towards this end, we first develop a comprehensive constitutive model able to capture all relevant nonlinear deformation physics of transient, reverse, and steady-state creep. The determination of the single set of representative material parameters is then achieved by framing the calibration process as an optimization problem, for which the global Particle Swarm Optimization algorithm is employed. Dynamic data integration is achieved by a multi-step calibration strategy for a situation where experiments are included one at a time, as they become available. Additionally, our calibration strategy is made flexible to account for mild heterogeneity between rock samples, resulting in a single set of parameters that is representative of the deformation datasets. As a rigorous mathematical analysis for the presented method and the lack of relevant experimental datasets, we consider a wide range of synthetic experimental data, inspired by the existing sparse relevant data in the literature. The results of our performance analyses show that the proposed calibration strategy is robust. Moreover, the model accuracy becomes increasingly better as more data is included for calibration.
	\end{abstract}
	
	
\end{frontmatter}

\section{Introduction}
Solution-mined salt caverns have been used for storage of hydrocarbons  \cite{koenig1994preparing,tarkowski2021storage,allen1972eminence,gentry1963storage}, compressed air \cite{crotogino2001huntorf}, and even feed-stock hydrogen mainly for chemical industry \cite{caglayan2020technical,tarkowski2019underground}. More recently, salt caverns are being considered as viable options for large-scale (e.g., in the orders of TWh) storage of renewable hydrogen to support energy transition  \cite{hashemi2021pore}. However, due to the intermittent nature of renewable sources and variable consumption rates, injections and productions of renewable hydrogen are expected to occur at much more unpredictable patterns than the current feed-stock-based storage systems. This can result in fast pressure fluctuations inside the caverns, which naturally increases the associated uncertainties related to the mechanical stability of the caverns. To perform stability analyses, selecting an appropriate constitutive model is the first crucial step. The constitutive model, used in the numerical simulation framework, is expected to allow for accurate prediction of the mechanical behavior of salt rock \cite{aubertin1999rate}. Constitutive models entail many parameters, which are expected to be tuned for a specific scenario. Therefore, an equally important step is the calibration of these material parameters such that the combination of the appropriate model and parameters leads to reliable safety assessments. Model calibration becomes especially challenging when many experimental data sets are available, for which the determination of a  single representative set of  parameters is required. This particular point is the main focus of this study.

Salt rocks entail a complex mechanical behavior, with different yet co-existing time-dependent deformation mechanisms \cite{cristescu1998time}. Capturing all these complexities in a single constitutive model is not a trivial task. In this category, the Hou/Lux \cite{hou2003mechanical} and CDM \cite{hampel2012cdm} models are developed for both transient and steady-state creep. In addition, they are also developed for tertiary creep in the dilatancy zone and damage healing in the compressibility zone. Conversely, other models disregard the compressibility/dilatancy regions and focus on specific deformation mechanisms, such as transient, steady-state, and tertiary creep \cite{heusermann2003nonlinear,zhou2011creep,jiang2013extended,ma2013new,deng2020viscoelastic,firme2018enhanced,firme2019salt}. However, as pointed out in the literature \cite{dusseault2004sequestration}, full constitutive models are not always necessary depending on the specific application and the stress regimes. For instance, transient creep might be disregarded for disposal operations, since internal pressure is expected to be fairly stable. However, this is not the case for hydrogen storage operations, where fast cyclic injections and productions are expected to take place. Although some researchers consider transient creep as a viscoelastic process \cite{zhou2011creep,heusermann2003nonlinear,deng2020viscoelastic}, others develop constitutive models that assume time-dependent inelastic deformations \cite{desai1987constitutive,jin1998elastic,hou2003mechanical,khaledi2016stability}. When rapid pressure swings are expected to occur, some studies in the literature also include reverse creep in their modeling framework \cite{munson1993extension,aubertin1999rate}.

The complexity of salt rock mechanics is often reflected in the constitutive models, which tend to depend on many material parameters. As an example, the viscoplastic part of the Hou/Lux-ODS and Hou/Lux-MDS \cite{hou2003mechanical} models depend on 11 and 18 material parameters, respectively. This makes the calibration strategy a very challenging task, which also requires many experimental data sets to characterize the material behavior. The viscoplastic model of Desai, for instance, requires a set of at least six experiments for determining all material parameters \cite{desai1987constitutive}. An additional challenge appears by recognizing the potential inherent differences between samples and difficulties in experimental controls \cite{robson2024calibration}. In other words, different material parameters might be found for the same batch of samples, thus requiring many sets of experiments for an ideal calibration procedure. In this case, a set of material parameters that is representative of all experiments should be pursued through the calibration process. Moreover, it should be noted that performing creep experiments in salt rocks is a time-consuming task \cite{yang1999experimental,ozcsen2014measurement,berest2015very}, which means that the size of the dataset increases at a low pace. Therefore, it would be convenient to perform partial calibrations with the experimental data currently available and include new experiments as they become available. During this process, the quality of the model results should increase as more experiments are used for calibration, and it should stabilize when a sufficient number of experiments is reached. 

There are different approaches for determining the material parameters of a constitutive model, some of which are complementary. As pointed out in \cite{anandarajah1991computer}, some material parameters, such as Young's modulus and Poisson's ratio, can be determined by well-defined procedures, while others are obtained either by manual (trial and error) or automated procedures, with the latter being preferred in face of a large number of parameters. A common approach for automated calibration consists of defining an appropriate objective function to be minimized in an optimization process. This is called direct calibration \cite{tolk2017advances}, and although some researchers choose gradient-based algorithms \cite{anandarajah1991computer}, meta-heuristic optimization approaches \cite{robson2024calibration,schulte2023machine} are more common choices as they are more likely to avoid local minima. The direct calibration approach treats the constitutive model as a black-box, so the optimization process is agnostic to the field of application. Nevertheless, such an approach is not found in the literature for constitutive models related to salt rocks. Furthermore, a procedure for incrementally including new experiments in the calibration process is also not reported in the literature.    

In this context, the present work proposes a calibration strategy for a salt rock constitutive model that includes deformation mechanisms relevant to cyclic operations in salt caverns, such as transient, steady-state, and reverse creep. The calibration of the material parameters related to each deformation mechanism is presented. A multi-step direct calibration procedure is developed to include one experiment at a time by properly adjusting the objective (loss) function and solving it as a multi-objective function optimization problem. Moreover, a regularization term is added to the loss function to roughly favor the same fitting quality for all experiments. The Particle Swarm Optimization (PSO) algorithm is employed for solving the optimization problems. Moreover, we investigate a situation where each experiment is performed on salt samples with slightly different material properties. As discussed before, the goal is to obtain a single set of representative material parameters. For this purpose, synthetic experiments are employed, such that quality assessment of the proposed calibration strategy is possible. Finally, a sensibility analysis is carried out to investigate the influence of some material parameters and to clarify the possible difficulties when performing the optimization process.

The remainder of this paper is organized as follows. Section \ref{sec:lab_experiment} presents experimental results obtained from a triaxial test on a salt rock. The experimental results serve as a guide for the choice of our constitutive model, which is presented in Section \ref{sec:model}. Once the model is defined, the calibration strategy adopted in this work is discussed in Section \ref{sec:calibration}. In Section \ref{sec:results}, the model validation against an experimental result is presented, and the calibration strategy is tested and investigated in different situations. Finally, Section \ref{sec:conclusion} closes our presentation.

\section{Experimental methodology and results}
\label{sec:lab_experiment}
To give additional insights into the mechanical behavior of salt rock and to provide a benchmark for the constitutive model employed in this work, we briefly discuss in this section a triaxial test performed on a salt rock. The experiment was performed on the Z3 Leine rock salt provided by the Institut fur Gebrismechanik GMBH (IFG), Germany. The Z3 Leine rock salt originated from the Bernburg mine in Germany, it is very homogeneous ($>98\%$ NaCl), colorless, and no obvious layering is visible. The grain size is less than 10 mm and the core sample was prepared as a 1-inch diameter and 2-inch long cylindrical shape. Additionally, the sample was stored inside an airtight bag and kept in a watertight container.

\begin{figure}[!ht]
	\centering
	\includegraphics[scale=0.65]{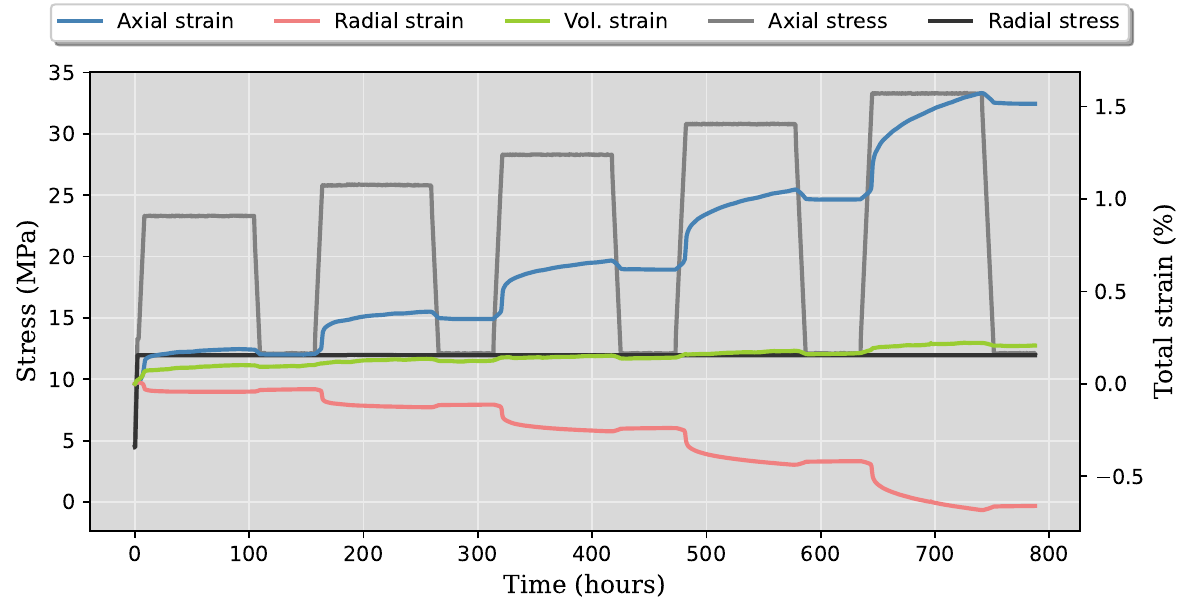}
	\caption{Experimental results obtained from a triaxial test performed on the Z3 Leine salt rock sample.}
    \label{fig:paper_lab_exp}
\end{figure}

We show here the methods and results of a cyclic loading test on the Z3 Leine rock salt. The experimental data is generated as testing data to assess the ability of our constitutive computer model to describe salt behavior under cyclic loading conditions. The experiment was carried out in a standard triaxial cell operating at room temperature (21°C). The choice for room temperature testing was made to keep the experiment as simple as possible. The triaxial cell was built in-house (Energy Transition Campus Amsterdam, ETCA) with an axial and radial stress limit of 95 MPa. The axial and radial displacements were measured during the test. We first loaded the cylindrical sample isostatically to 12.9 MPa and let the samples consolidate until the measured axial and radial strain equilibrated. 
After equilibrium, the radial stress was kept at 12.9 MPa for the duration of the experiment, whereas the axial stress was increased to 23.5 MPa (at 2.0 MPa$/$hr) and kept at this stress for 4 days and thereafter, decreased back down to 12.9 MPa (at 2.0 MPa$/$hr) followed by a consolidation for another 2 days. The 4 days of high axial stress and 2 days of low axial stress were repeated for other four axial stress steps, namely, 26.0/12.9 MPa, 27.5/12.9 MPa, 31.0/12.9 MPa, 33.5/12.9 MPa (see Fig. \ref{fig:paper_lab_exp}). Figure \ref{fig:paper_lab_exp} also shows axial and radial strains, both measured during the experiment, and volumetric strain (calculated) versus time. In this case, all five differential stress steps show axial shortening and volumetric shrinkage, whereas the radial strain shows dilation in all differential stress steps.

\begin{figure}[!ht]
	\centering
	\includegraphics[scale=0.60]{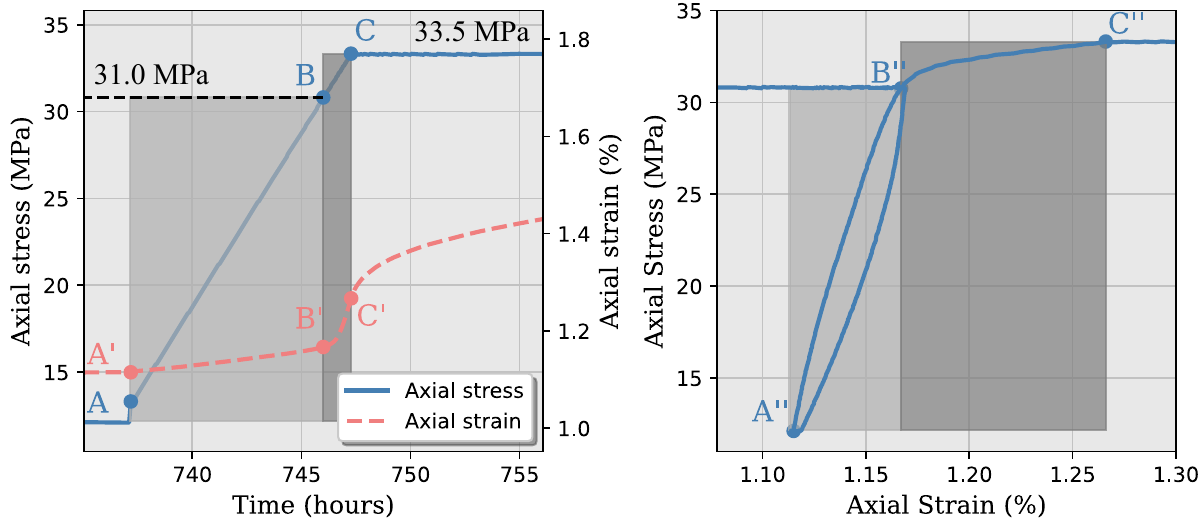}
	\caption{Zoomed-in view of a reloading step of the experiment that allows for identification of both viscoplastic and viscoelastic deformations.}
    \label{fig:further_discussion}
\end{figure}

A closer look at this experiment can provide valuable insights into the modeling of salt mechanics. For this purpose, the left graph of Fig. \ref{fig:further_discussion} shows a zoomed view of the last loading stage (i.e. from 12.9 to 33.5 MPa). Notice that the black dashed line indicates the previous maximum stress of 31.0 MPa. When the axial stress goes from point A to point B, the corresponding strain goes from point A’ to B’, as indicated by the dashed red curve. As can be seen, the strain increment is relatively small along the stress path A-B. However, when the axial stress exceeds the maximum stress ever applied to the sample during the experiment, which at this point is 31.0 MPa, an abrupt increase in strain is observed from point B’ to C’. This suggests that an additional inelastic strain has been triggered during the stress path B-C. This can be also concluded from the right graph of Fig. \ref{fig:further_discussion}, which shows the corresponding stress-strain relation during the loading path A-B-C. To facilitate the visualization, notice that the light and dark gray areas in the two graphs of Fig. \ref{fig:further_discussion} correspond. The right graph shows that from point A” to B”, the salt sample behaves as a (visco)elastic material. However, from point B” to C” there is a large increase in strain for a relatively small stress increment. This is a characteristic of plastic (or viscoplastic) deformation, and it can be observed in every unloading/reloading cycle during the experiment.

Finally, the right graph of Fig. \ref{fig:further_discussion} shows a hysteretic effect during the unload/reload path. The unload path goes from point B” to point A” and the load path goes from point A” back to B”. The fact that these two paths are not the same is an indication of the hysteretic effect. It is well-known that viscoelastic materials present this type of behavior during loading/unloading cycles. This indicates that, in addition to viscoplasticity, the salt rock also presents a viscoelastic behavior. This hysteretic effect is actually a manifestation of the well-known reverse creep.

\section{Model formulation}
\label{sec:model}
Following the infinitesimal deformation assumption, the total strain can be decomposed into different independent contributions by different mechanisms. The first contribution accounts for the instantaneous elastic response. Reverse creep, which is observed after an unloading step, is interpreted as a time-dependent recoverable strain (i.e. viscoelastic) and is described by a Kelvin-Voigt element \cite{munson1993extension}. Transient creep, on the other hand, is regarded as a time-dependent inelastic deformation (i.e., viscoplastic) \cite{khaledi2016stability}. Finally, steady-state dislocation creep is captured by a dashpot element following a power-law function. In this manner, the total strain can be written as
\begin{equation}
	\pmb{\varepsilon} = \pmb{\varepsilon}_{e} + \pmb{\varepsilon}_{ve} + \pmb{\varepsilon}_{vp} + \pmb{\varepsilon}_{cr},
	\label{eq:total_strain_rate}
\end{equation}
where $\pmb{\varepsilon}_{e}$, $\pmb{\varepsilon}_{ve}$, $\pmb{\varepsilon}_{vp}$ and $\pmb{\varepsilon}_{cr}$ denote the elastic, viscoelastic, viscoplastic and steady-state creep strain tensors, respectively. Next, the specific models for each of these contributions are described.

\subsection{Elastic strain}
For convenience, we define a function of Poisson's ratio, $\nu$, as $\mathbb{C} = \mathbb{C}(\nu)$, such that $\mathbb{C}:\mathbb{R} \rightarrow \mathbb{R}^{3\times 3\times 3\times 3}$ is a 4th-order tensor with elements given by
\begin{equation}
	\mathbb{C}_{ijkl}(\nu) = \frac{\nu}{(1+\nu)(1-2\nu)} \delta_{ij} \delta_{kl} + \frac{1}{2(1+\nu)} ( \delta_{ik} \delta_{jl} + \delta_{il} \delta_{jk} ).
	\label{eq:C_poisson}
\end{equation}

The time-independent elastic strain tensor at any time $t_i$ can be calculated with the linear elasticity, assuming isotropic properties, i.e.,
\begin{equation}
	\pmb{\varepsilon}_e(t_i) = \mathbb{C}_1^{-1} : \pmb{\sigma}(t_i),
	\label{eq:eps_e}
\end{equation}
where $\mathbb{C}_1$ is the constitutive 4th-order tensor, defined as
\begin{equation}
	\mathbb{C}_1 = E_1 \mathbb{C}(\nu_1).
\end{equation}

\noindent Here, $E_1$ and $\nu_1$ denote the Young's modulus and Poisson's ratio associated to the elastic deformation, respectively.

\subsection{Viscoelastic strain}
\label{subsec:viscoelastic}
The solution for the time-dependent elastic (i.e. viscoelastic) strain, represented by a Kelvin-Voigt element, for continuous stress functions has been studied in the literature \cite{argyris1991constitutive}. However, experimental data sets are discrete, not continuous. Therefore, in this work, we present a  solution strategy for discrete stress conditions. 

\begin{figure}[!b]
	\centering
	\includegraphics[scale=0.95]{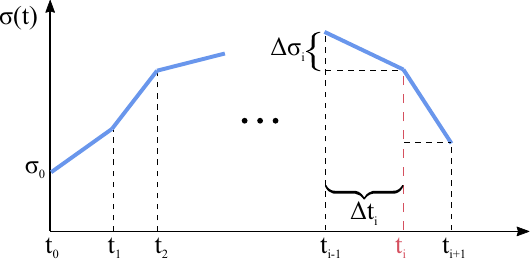}
	\caption{Discrete stress path in time.}
    \label{fig:viscoelastic_solution}
\end{figure}

Towards this end, consider a discrete stress function as depicted in Fig. \ref{fig:viscoelastic_solution}. The viscoelastic strain (i.e., for the Kelvin-Voigt element) subjected to this discrete stress condition reads
\begin{equation}
    \pmb{\varepsilon}_{ve}(t_i) = \mathbb{C}_2^{-1}(t_i) : \pmb{\sigma}_0
    + \sum_{j=0}^{i-1} \mathbb{C}_2^{-1}(t_i - t_j) : \Delta \pmb{\sigma}_{j+1},
    \label{eq:eps_ve}
\end{equation}
where
\begin{equation}
    \mathbb{C}_2^{-1}(t) = \frac{1}{E_2} \left( 1 - e^{-\frac{E_2}{\eta_2}t} \right)  \mathbb{C}^{-1}(\nu_2)
\end{equation}
holds. Here, $\pmb{\sigma}_0$ represents the stress tensor at $t=t_0$ and  $\mathbb{C}(\nu_2)$ is defined by Eq. \eqref{eq:C_poisson}. Moreover, $E_2$ and $\nu_2$ denote the Young's modulus and Poisson's ratio of the spring in the Kelvin-Voigt element, and $\eta_2$ is the dashpot viscosity of the same element.

\subsection{Viscoplastic strain}
\label{subsec:viscoplastic}
Following Perzyna's theory \cite{desai1987constitutive,khaledi2016stability}, the viscoplastic strain rate is expressed as
\begin{equation}
    \dot{\pmb{\varepsilon}}_{vp} = \mu_1 \left\langle \dfrac{ F_{vp} }{F_0} \right\rangle^{N_1} \dfrac{\partial Q_{vp}}{\partial \pmb{\sigma}},
    \label{eq:eps_vp_rate}
\end{equation}
where $\mu_1$ and $N_1$ are material parameters, $F_0$ is a reference value, and the operator $\left\langle \cdot \right\rangle$ denotes the ramp function. The yield function \cite{desai1987constitutive} is given by 
\begin{equation}
    F_{vp}(\alpha, \pmb{\sigma}) = J_2 - (-\alpha I_1^{n_1} + \gamma I_1^2) \left[ \exp{(\beta_1 I_1)} + \beta \dfrac{\sqrt{27} J_3}{2 \sqrt{J_2^3}} \right]^m \label{eq:F_vp}
\end{equation}
in which $n_1$, $\gamma$, $\beta_1$, $\beta$ and $m$ are material parameters. In addition, an associative viscoplastic deformation mechanism is considered, implying that the potential function $Q_{vp}$ in Eq. \eqref{eq:eps_vp_rate} is equal to the yield function $F_{vp}$ of Eq. \eqref{eq:F_vp}.

Moreover, denoting the deviatoric stress tensor as $\mathbf{s}$, the stress invariants in Eq. \eqref{eq:F_vp} can be written as
\begin{equation}
	I_1 = \text{tr}(\pmb{\sigma}) + \sigma_t,
	\quad
	J_2 = \dfrac{1}{2} \mathbf{s} : \mathbf{s},
	\quad \text{and} \quad
	J_3 = \text{det}(\pmb{\sigma}),
\end{equation}
where $\sigma_t$ is the tensile strength of the rock.

The hardening parameter in Eq. \eqref{eq:F_vp} characterizes the behavior of the yield function. The hardening rule adopted in this work is
\begin{equation}
	\alpha = \dfrac{a_1}{\left[ \left(\dfrac{a_1}{\alpha_0}\right)^\frac{1}{\eta} + \xi \right]^\eta},
	\label{eq:hardening}
\end{equation}
where $\eta$, $a_1$, $\alpha_0$ are material parameters. In particular, $\alpha_0$ denotes the initial configuration of the yield surface. Additionally, $\xi$ is the accumulated viscoplastic strain, which is given by
\begin{equation}
	\xi = \int_{t_0}^t \sqrt{\dot{\pmb{\varepsilon}}_{vp} : \dot{\pmb{\varepsilon}}_{vp}} \text{d}t.
	\label{eq:qsi}
\end{equation}

\begin{figure}[!ht]
	\centering
	\includegraphics[scale=0.65]{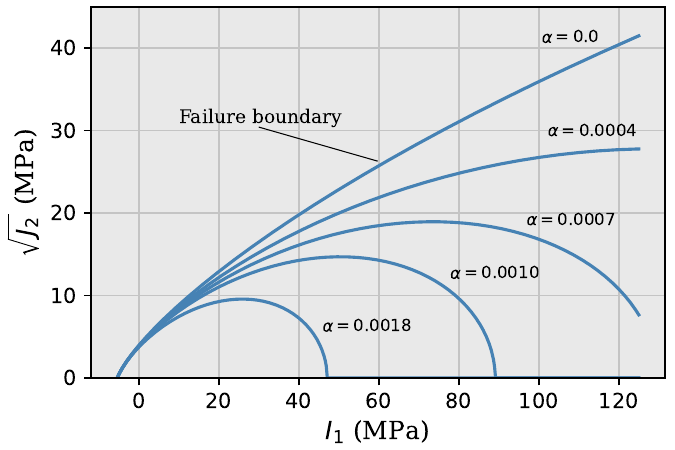}
	\caption{Yield function for the viscoplastic model for different hardening parameters.}
    \label{fig:F_vp}
\end{figure}

\noindent Note that, before any viscoplastic deformation takes place, Eq. \eqref{eq:qsi} results in $\xi=0$ and Eq. \eqref{eq:hardening} leads to $\alpha = \alpha_0$, as expected. Furthermore, $\alpha\rightarrow0$ when $\xi\rightarrow\infty$, which causes the yield surface to assume the position of the short-term failure boundary. Figure \ref{fig:F_vp} shows the yield surface (i.e. $F_{vp}(\alpha, \pmb{\sigma})=0$) for different values of hardening parameter $\alpha$.

Once the viscoplastic strain rate is obtained, the viscoplastic strain at time $t_i$ is obtained by performing a time integration and using backward Euler scheme, i.e.,
\begin{equation}
	\pmb{\varepsilon}_{vp}(t_i) = \pmb{\varepsilon}_{vp}(t_{i-1}) + \Delta t \dot{\pmb{\varepsilon}}_{vp}(t_i),
\end{equation}
in which $\Delta t = t_i - t_{i-1}$. Note that the viscoplastic strain rate is also computed at $t_i$ due to the implicit formulation.

\subsection{Dislocation creep strain}
In this study, we neglect pressure solution creep and consider dislocation creep only, as most of the available data in the literature are conducted under deviatoric stresses exceeding 5 MPa \cite{kumar2022influence}. Dislocation creep is often modeled as using a power law \cite{ramesh2021geomechanical}, which reads
\begin{equation}
	\dot{\pmb{\varepsilon}}_{cr} = A \exp\left( - \dfrac{Q}{RT} \right) q^{n-1} \mathbf{s},
	\label{eq:eps_cr_rate}
\end{equation}
where $A$ and $n$ are material parameters, $Q$ is the activation energy, $T$ is temperature and $R$ is the universal gas constant ($R=8.32$ JK$^{-1}$mol$^{-1}$).

\noindent Similar to the viscoplastic strain, the creep strain at time $t_i$ is obtained by
\begin{equation}
	\pmb{\varepsilon}_{cr}(t_i) = \pmb{\varepsilon}_{cr}(t_{i-1}) + \Delta t \dot{\pmb{\varepsilon}}_{cr}(t_i),
	\label{eq:eps_cr}
\end{equation}
where implicit time integration is also employed.

\section{The model calibration strategy based on ensemble of data sets}
\label{sec:calibration}
In this section, we present the developed calibration strategy adopted to determine the material parameters of the multi-physics constitutive model. As summarized in Table \ref{tab:calibration_params}, the multi-physics model entails 19 material parameters\footnote{The variable $\alpha_0$ defines the initial position of the yield surface. Therefore, it is not strictly a material parameter but rather an initial condition. Nevertheless, for the purposes of this work, $\alpha_0$ is also regarded as a material parameter.} in total. Any manual strategy to fit them would be certainly inapplicable. Moreover, performing optimization in a 19th-dimensional space is cumbersome and prone to suffer from the local optimum traps or too many possibilities to match the given data set. As a remedy to this significant challenge, in this study, we propose a strategy to adjust the material parameters separately in three different groups as follows. The first group, i.e., group 1, comprises the parameters associated with the dislocation creep model. The second group, i.e., group 2, is the union of the material parameters representing both elastic and viscoelastic models. Finally, the parameters related to the viscoplastic model are considered in the third group, i.e., group 3. The sequence of calibration goes from group 1 to 3. A brief discussion of the calibration steps of each group is provided below.

\begin{table}[!ht]
	\centering
	\caption{Material parameters to be calibrated.}
	\begin{tabular}{clc}
		\hline
		\multicolumn{1}{l}{\textbf{Group}} & \textbf{Model} & \textbf{Parameters}                    \\ \hline
		1                                                   & Dislocation creep               & $A$, $n$                                                \\ \hline
		\multirow{2}{*}{2}                                  & Elastic                         & $E_1$, $\nu_1$                                            \\ \cline{2-3} 
		                                                    & Viscoelastic                    & $E_2$, $\nu_2$, $\eta_2$                                         \\ \hline
		\multirow{2}{*}{3}                                  & \multirow{2}{*}{Viscoplastic}   & $\mu_1$, $\eta$, $N_1$, $n_1$, $\beta$, $m$             \\
		                                                    &                                 & $a_1$, $\beta_1$, $\gamma$, $\alpha_0$, $k$, $\sigma_t$ \\ \hline
	\end{tabular}
	\label{tab:calibration_params}
\end{table}

\subsection{Group 1: Dislocation creep model}
\begin{figure}[!ht]
	\centering
	\includegraphics[scale=0.65]{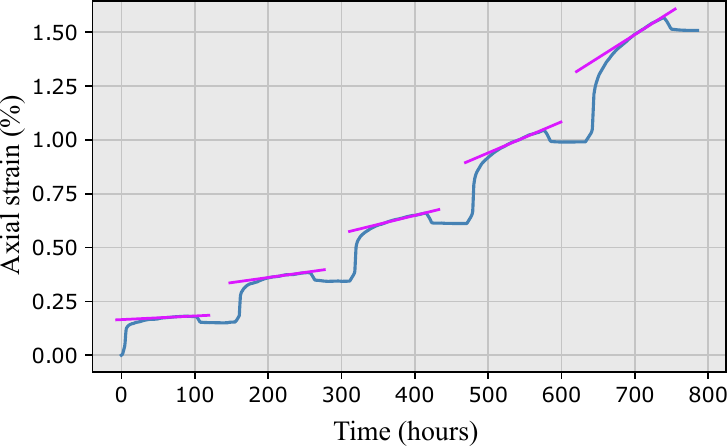}
	\caption{Steady state creep rates (slopes) represented by the magenta lines.}
    \label{fig:fit_creep}
\end{figure}

As expressed in Eq. \eqref{eq:eps_cr_rate}, the creep model depends on 2 material parameters: $A$ and $n$. Typical values of $n$ for dislocation creep range between 3 to 7 \cite{hunfeld2022influence}. Based on this information, we adjust parameters $A$ and $n$ such that the slopes of the creep strain curve match those from the experimental data. For example, Figure \eqref{fig:fit_creep} shows the axial strain measured from the experimental results and the target slopes (stationary creep strain rates) represented by the magenta lines. For this step, only the solution provided by Eq. \eqref{eq:eps_cr} is necessary, since only the slopes of the curves are required.

The material parameters found for the dislocation creep model are summarized in Table \ref{tab:params_group_1}.

\begin{table}[!b]
	\centering
	\caption{Material parameters for Group 1.}
	\begin{tabular}{ccc}
		\textbf{Parameter} & \textbf{Value}           & \textbf{Unit}  \\ \hline
		$n$   	& 3.0             		& -    \\
		$A$     & $1.1\times 10^{-21}$	& Pa$^{-n}$s$^{-1}$
	\end{tabular}
	\label{tab:params_group_1}
\end{table}

\subsection{Group 2: Elastic and viscoelastic models}
\begin{figure}[!ht]
	\centering
	\includegraphics[scale=0.65]{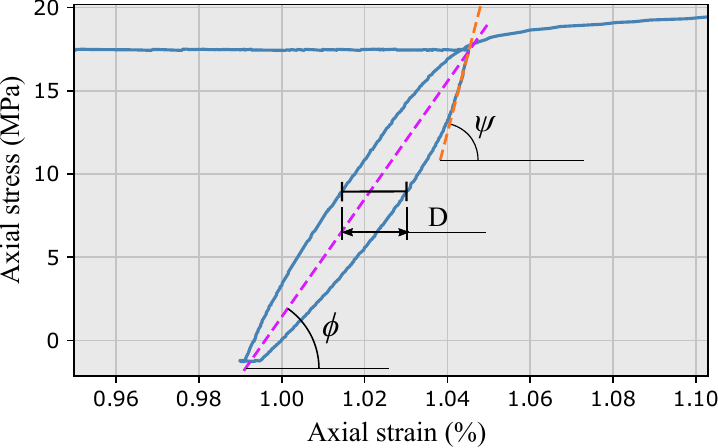}
	\caption{Features ($\phi$ and $D$) observed when adjusting the elastic and viscoelastic parameters.}
    \label{fig:fit_viscoelastic}
\end{figure}

For the elastic model, the material parameters are the Young modulus and Poisson's ratio. The latter can be determined by the ratio between radial and axial strains. Additionally, is it assumed that the Poisson's ratio for elastic and viscoelastic elements are the same, i.e., $\nu_1=\nu_2$. The Young's modulus, $E_1$, can be obtained by the slope $\Psi$ of the unloading/loading path, as illustrated in Fig. \ref{fig:fit_viscoelastic}. Finally, the values of $E_2$ and $\eta_2$ are adjusted all together to match the distance $D$ and the angle $\phi$ of the average unloading/loading slope.

For the elastic and viscoelastic contributions of the constitutive model, the material parameters obtained based on the above described procedure are shown in Table \ref{tab:params_group_2}.

\begin{table}[!ht]
	\centering
	\caption{Material parameters for Group 2.}
	\begin{tabular}{ccc}
		\textbf{Parameter} & \textbf{Value}           & \textbf{Unit}  \\ \hline
		$E_1$     & 102             & GPa   \\
		$\nu_1$   & 0.32            & -     \\
		$E_2$     & 42              & GPa   \\
		$\nu_2$    & 0.32            & -     \\
		$\eta_2$  & $2.5\times10^5$ & GPa.s
	\end{tabular}
	\label{tab:params_group_2}
\end{table}

\subsection{Group 3: Viscoplastic model}
As summarized in Table \ref{tab:calibration_params}, the viscoplastic model requires the definition of 12 parameters. Although different types of salt rocks and the presence of impurities may produce different values for these material properties, typical values for salt rocks can be found in the literature \cite{khaledi2016stability, desai1987constitutive}. The parameter $\gamma$ is associated with the short-term failure boundary, whereas $\beta$, $\beta_1$ and $m$ account for the influence of Lode's angle. The position of the dilatancy boundary can be adjusted mainly through parameters $\beta_1$ and $n_1$. Also, $F_0$ is a reference value and $k$ is regarded as zero, meaning that an associative flow rule is adopted in this work. These parameters are summarized in Table \ref{tab:params_group_3}.

The remaining parameters $\mu_1$ and $N_1$ relate to the rate-dependent behavior of the salt rock and should be calibrated for specific samples. Similarly, the hardening rule described by Eq. \eqref{eq:hardening} depends on $a_1$ and $\eta$, which should also be calibrated. Additionally, the initial position of the yield surface is defined by the parameter $\alpha_0$, which depends on the stress history of the rock sample. The definition of $\alpha_0$ is very important for the accuracy of the predictions, in the presence of viscoplasticity. 

Based on the above discussion, from the 12 material parameters associated with the viscoplastic model, only $\mu_1$, $N_1$, $a_1$, $\eta$ and $\alpha_0$ are to be determined. The remaining parameters are taken from typical values found in the literature. Next, the strategy developed in this work for obtaining these sets of parameters is described.

\begin{table}[!ht]
    \centering
    \caption{Fixed material parameters for the viscoplastic element.}
    \begin{tabular}{ccc}
        \textbf{Parameter} & \textbf{Value}      & \textbf{Unit} \\ \hline
        $n_1$                & $3$                 & $-$           \\
        $\beta_1$          & $4.8\times 10^{-3}$ & MPa$^{-1}$    \\
        $\beta$            & $0.995$             & $-$           \\
        $m$                & $-0.5$              & $-$           \\
        $\gamma$           & $0.095$             & $-$           \\
        $F_0$              & $1.0$               & MPa$^2$       \\    
        $k$                & $0.0$               & $-$       \\    
    \end{tabular}
    \label{tab:params_group_3}
\end{table}

\subsection{Optimization strategy}
As shown in the previous subsections, the constitutive model depends on 19 material parameters. Let us denote the entire set of material parameters as $\textbf{m} \in \mathbb{R}^{d\times 1}$, where $d=19$. Furthermore, for the purpose of the optimization procedure, let us split $\textbf{m}$ in two subsets: $\textbf{w} \in \mathbb{R}^{w \times 1}$, denoting the material parameters to be optimized; and $\textbf{k} \in \mathbb{R}^{k \times 1}$ comprising the material parameters that are fixed. Mathematically, $\textbf{m} = \textbf{w} \cup \textbf{k}$, which implies that $w+k=d$. As discussed before, the material parameters to be optimized are
\begin{equation}
	\textbf{w} = 
	\begin{bmatrix}
		\mu_1 &
		N_1 &
		a_1 &
		\eta &
		\alpha_0
	\end{bmatrix}^T,
\end{equation}

\noindent whereas $\textbf{k}$ comprises all the remaining material parameters, which are fixed (Table \ref{tab:params_group_3}). Note that the set of material parameters to be optimized ($\mathbf{w}$) belong exclusively to the viscoplastic contribution of the constitutive model.

The strain tensor for a certain experiment $i$ is then obtained by
\begin{equation}
	\pmb{\varepsilon}_i(\textbf{w}, t) = f(\textbf{w}, \textbf{k}, \pmb{\sigma}_i(t))
	\quad
	\text{for} \hspace{1.5mm} i \in [1, \cdots, N],
\end{equation}

\noindent where $N$ is the number of experiments and $\pmb{\sigma}_i(t)$ is the specific stress condition applied to experiment number $i$.

For each experiment $i$, we define a loss function $F_i(\textbf{w})$ by taking the Mean Absolute Percentage Error (MAPE) of the axial ($\varepsilon_{a}$) and radial ($\varepsilon_{r}$) strains, i.e.,
\begin{equation}
	F_i(\textbf{w}) = \dfrac{1}{N_t} \sum_{j=1}^{N_t} \left( \left|\left| \frac{\varepsilon_{a,i}(\textbf{w}, t_j) - \varepsilon_a^*(t_j)}{\varepsilon_a^*(t_j)} \right|\right| + \left|\left| \frac{\varepsilon_{r,i}(\textbf{w}, t_j) - \varepsilon_r^*(t_j)}{\varepsilon_r^*(t_j)} \right|\right| \right),
    \label{eq:loss_F}
\end{equation}

\noindent where $||\cdot||$ represents the absolute value, $N_t$ is the number of time steps in experiment $i$, and $\pmb{\varepsilon}^*(t_j)$ is the true deformation measured at time $t_j$. The MAPE metric represents an average of the percentage error at each time $t_i$. By considering the percentage error, the same importance is given to the errors in both the beginning and the end of the experiment.

When more than one experiment is considered, say $i$ experiments, the loss functions associated with each experiment are combined by taking the Mean Squared Error (MSE) plus a regularization term, i.e.,
\begin{equation}
    L_i(\textbf{w}) = \dfrac{1}{i} \sum_{j=1}^i F_j^2(\textbf{w}) + \phi \sigma^2,
    \label{eq:loss_L}
\end{equation}

\noindent in which $\phi$ is the regularization factor and $\sigma^2$ is the variance of the errors in all $i$ experiments. The incorporation of a regularization term is designed to mitigate the risk of the optimization process becoming overly specialized to a single experiment, thereby compromising its generalizability to other experiments. This is further discussed in subsection \ref{subsec:loss_reg}.

Every time a new experiment is added, an optimization problem is solved to obtain the best set of material properties for that particular experiment only. For a given experiment $\#i$, for example, this is achieved by minimizing the corresponding loss function $F_i$. In other words, the set of material parameters resulting from this process is defined as

\begin{equation}
	\textbf{w}_i^F \approx \text{argmin} \hspace{0.5mm} F_i(\textbf{w}) \quad \forall \hspace{1.5mm} \textbf{w} \in \mathbb{P},
    \label{eq:w_F_i}
\end{equation}

\noindent where $\mathbb{P} \subset \mathbb{R}^{w \times n}$ is the initial population distribution, with $n$ denoting the initial population (swarm) size. The approximate sign ``$\approx$'' is used in Eq. \ref{eq:w_F_i} to admit the possibility of obtaining sub-optimal or, at least, approximate solutions.

Once the single experiment optimization has been performed for experiment $\#i$, the set of parameters $\textbf{w}_i^F$ is added to the population of the next full optimization, which considers all experiments from 1 to $i$. That is,

\begin{equation}
	\textbf{w}_i^L \approx \text{argmin} \hspace{0.5mm} L_i(\textbf{w}) \quad \forall \hspace{1.5mm} \textbf{w} \in \mathbb{P} \cup \mathbb{F}_i \cup \mathbb{L}_i.
    \label{eq:w_L_i}
\end{equation}

\noindent Here, $\mathbb{F}_i$ and $\mathbb{L}_i$ represent the union of the solutions obtained from optimizing Equations \eqref{eq:w_F_i} and \eqref{eq:w_L_i}, respectively. In other words,

\begin{equation}
	\mathbb{F}_i = \bigcup_{j=1}^i \textbf{w}_j^F
	\quad \text{and} \quad
	\mathbb{L}_i = \bigcup_{j=1}^{i-1} \textbf{w}_j^L.
\end{equation}

\noindent Notice that the full optimization (Eq. \ref{eq:w_L_i}) is performed with the initial population plus the individuals contained in sets $\mathbb{F}_i$ and $\mathbb{L}_i$. This means that the set of parameters obtained from the previous full optimizations is also considered for the current full optimization. In this manner, the calibrations of subsequent experiments do not start from scratch, as they consider the results from previous calibrations. 

The complete procedure is described in Algorithm \ref{alg:alg_1}. Notice that $\mathbb{F}$ and $\mathbb{L}$ are initially empty sets. Moreover, when only one experiment is available ($i=1$), there is no full optimization, so $\mathbb{L}$ remains empty. After the second experiment has been added, full optimizations are performed and $\bar{\textbf{w}}_i^L$ are added to $\mathbb{L}$. The algorithm loops over $N$ experiments, including them one by one to the calibration process.

\begin{algorithm}
	\caption{Strategy for simultaneous calibration of a set of experiments.}
	\begin{algorithmic}
	    \State $\mathbb{F} = \emptyset$
	    \State $\mathbb{L} = \emptyset$
	    \For {$i\leftarrow 1 \cdots N$}
	        \State $\textbf{w}_i^F \approx \text{argmin} \hspace{0.5mm} F_i(\textbf{w}) \quad \forall \hspace{1.5mm} \textbf{w} \in \mathbb{P}$
	        \State $\mathbb{F} \leftarrow \mathbb{F} \cup \textbf{w}_i^F$
	        \If {$i > 1$}
	            \State $\bar{\textbf{w}}_i^L \approx \text{argmin} \hspace{0.5mm} L_i(\textbf{w}) \quad \forall \hspace{1.5mm} \textbf{w} \in \mathbb{P} \cup \mathbb{F} \cup \mathbb{L}$
	            \State $\mathbb{L} \leftarrow \mathbb{L} \cup \textbf{w}_i^L$
	        \EndIf
	    \EndFor
	\end{algorithmic}
    \label{alg:alg_1}
\end{algorithm}

\subsubsection{Optimization algorithm}
The solutions of the optimization problems (Equations \ref{eq:w_F_i} and \ref{eq:w_L_i}) are obtained by the Particle Swarm Optimization (PSO) algorithm \cite{eberhart1995new,gad2022particle}. The PSO is a meta-heuristic optimization algorithm based on population intelligence. An initial population is created within the search space, and each member of the population represents a possible solution. Each individual is assigned an initial velocity so that they can move. For each time step (iteration), the velocity of a certain particle $i$ is computed as

\begin{equation}
    \mathbf{v}^{t+1}_i = \omega \mathbf{v}^t_i + c_1 \mathbf{r}_1 \left( \mathbf{p}_i^t - \mathbf{x}_i^t \right) + c_2 \mathbf{r}_2 \left( \mathbf{g}_i^t - \mathbf{x}_i^t \right),
    \label{eq:pso_params}
\end{equation}

\noindent where $\omega$, $c_1$ and $c_2$ are weights given to the previous velocity (inertia), particle's best position ($\mathbf{p}_i^t$) and swarm's best position ($\mathbf{g}_i^t$), respectively. Moreover, $\mathbf{r}_1$ and $\mathbf{r}_2$ are random vectors between 0 and 1. With this new velocity, the position of particle $i$ is updated by

\begin{equation}
    \mathbf{x}^{t+1}_i = \mathbf{x}^t_i + \mathbf{v}_i^{t+1}.
\end{equation}

Once all particles have their positions updated, the algorithm proceeds to the next iteration. This process is repeated until a stop criterion is reached.

\subsubsection{Multi-objective function optimization}
\label{subsec:loss_reg}
\begin{figure}[!ht]
	\centering
	\includegraphics[scale=0.35]{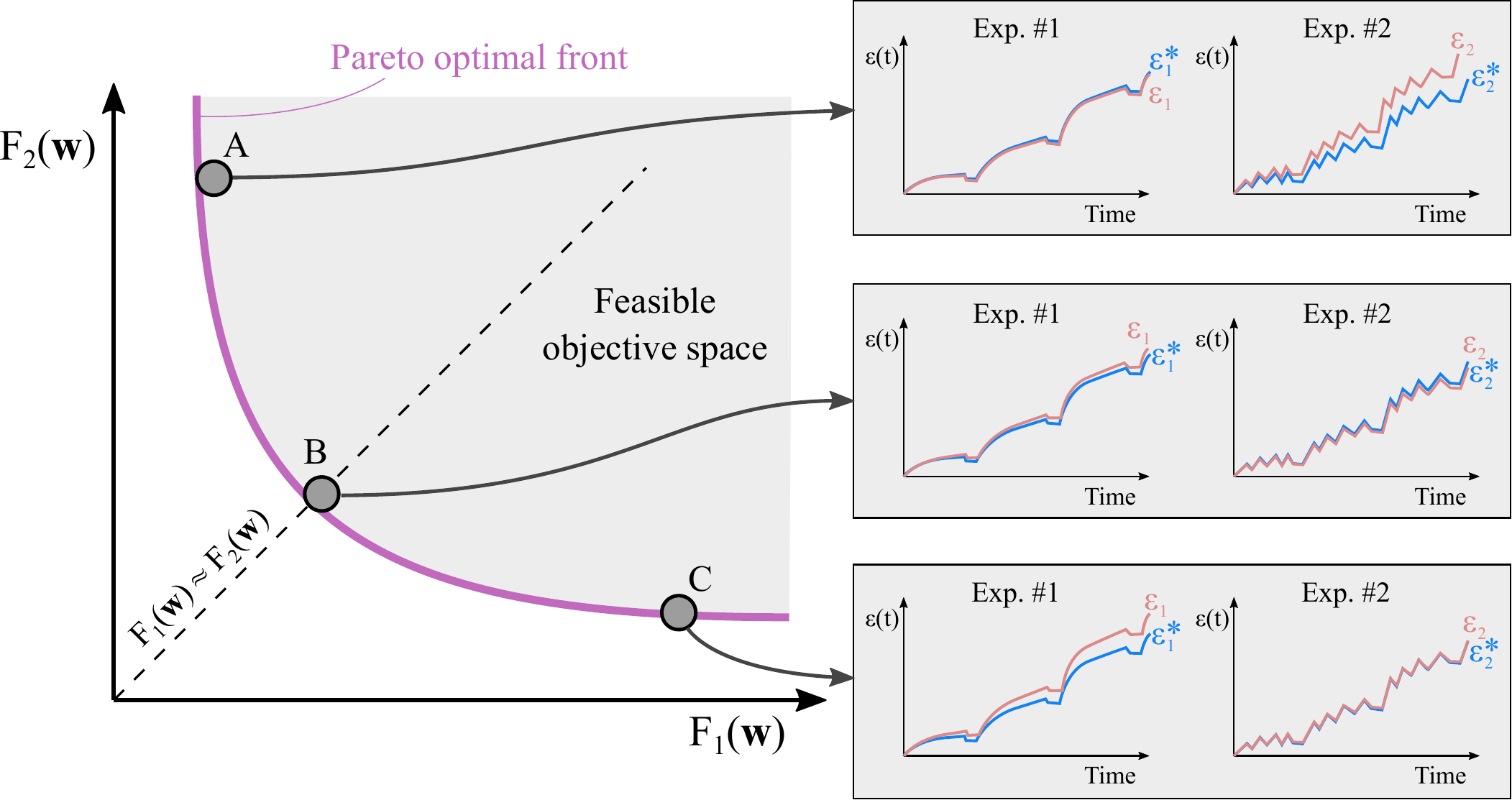}
	\caption{Multi-objective optimization and Pareto front. Experiments \#1 and \#2 in this figure are only meant for illustration purposes.}
    \label{fig:paretto}
\end{figure}

Solving an optimization problem for multiple experiments simultaneously (Eq. \ref{eq:w_L_i}) consists of multi-objective function optimization, since the goal is to minimize the loss functions $F_i$ for all experiments. This type of problem admits an infinite number of solutions that are not dominated by other feasible solutions, which means that they are all acceptable solutions to the minimization problem. This set of solutions is called the Pareto optimal front and is illustrated in Fig. \ref{fig:paretto} for a situation where only two experiments are considered. 

Although all solutions along the Pareto front are admissible, we notice that the solution at point A provides a good fit for experiment \#1 and a sub-optimal solution for experiment \#2. Similarly, point C provides a better fit for experiment \#2 than for \#1. On the other hand, point B provides a solution that is equally good for both experiments, since $F_1 \approx F_2$, and therefore should be preferred.

\begin{figure}[!ht]
	\centering
	\includegraphics[scale=0.7]{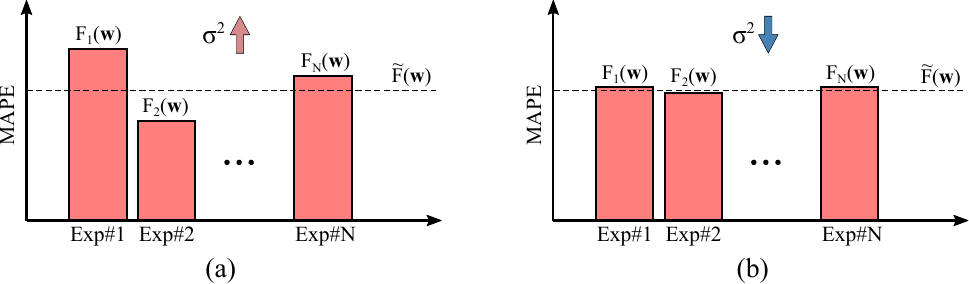}
	\caption{Variance of $F_i$. (a) High variance, (b) low variance.}
    \label{fig:variance}
\end{figure}

To ``encourage'' the solution to stay along the dashed line of Fig. \ref{fig:paretto} we penalize the loss function $L_i$ when $F_1$ $\cdots$ $F_N$ are too different from each other, that is, when the variance is high (see Fig. \ref{fig:variance}). Mathematically, this can be written as

\begin{equation}
    L_i(\mathbf{w})
    = \dfrac{1}{i} \sum_{j=1}^i F_j^2(\textbf{w}) + \phi \underbrace{ \dfrac{1}{i} \sum_{j=1}^i \left[ F_j(\mathbf{w}) - \tilde{F}(\mathbf{w}) \right] }_{\sigma^2}
    = \dfrac{1}{i} \sum_{j=1}^i F_j^2(\textbf{w}) + \phi \sigma^2,
    \label{eq:loss_reg}
\end{equation}

\noindent where $\tilde{F}(\mathbf{w})$ is the average error of all experiments, as indicated in Fig. \ref{fig:variance}, and $\phi$ is an arbitrary regularization factor. When variance is high (Fig. \ref{fig:variance}-a), the regularization term is large, thus penalizing the loss function $L_i$. When variance is close to zero (Fig. \ref{fig:variance}-b), which means that the fit quality for all experiments is about the same, the regularization term tends to vanish from Eq. \eqref{eq:loss_reg}.

\section{Results}
\label{sec:results}
In the first part of this section, we investigate the capabilities of the proposed constitutive model in describing the mechanical behavior of salt rocks observed in the laboratory. In the sequence, the optimization procedure proposed for the model calibration is explored. We first start by presenting the synthetic data employed throughout our analysis. Sensitivity analysis is performed to identify the impact of each material parameter on the model behavior. The PSO algorithm is first employed to calibrate each synthetic experiment individually. Then, the calibration strategy is tested for fitting the entire set of experiments as they become available, as described in Algorithm \ref{alg:alg_1}.

\subsection{Model validation}
\begin{figure}[!ht]
	\centering
	\includegraphics[scale=0.85]{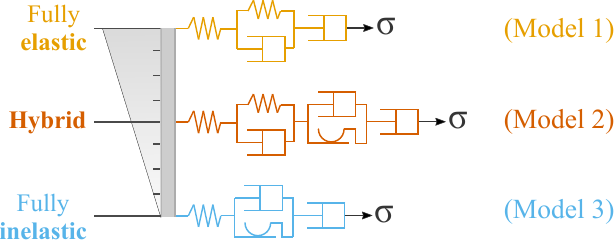}
	\caption{Models with different assumptions regarding the transient creep.}
    \label{fig:models_1_2_3}
\end{figure}

The primary purpose of this subsection is to show that the constitutive model presented in Section \ref{sec:model} is able to accurately capture the time-dependent behavior of salt rocks operating in the compressibility zone (i.e., no tertiary creep) and under cyclic loading conditions. Additionally, we intend to highlight the importance of each term considered in the proposed constitutive model. Of particular interest is the description of transient creep, as it is permanently present during cyclic operations. For this purpose, we consider three models with different characteristics regarding the transient creep description, as shown in Figure \ref{fig:models_1_2_3}. In \textit{Model 1}, the transient creep deformations are considered to be fully recoverable (elastic). They thus can be described by the Kelvin-Voigt element presented in subsection \ref{subsec:viscoelastic} (see Fig. \ref{fig:models_1_2_3}). On the other extreme, the hypothesis that all the deformation observed during the primary stage of creep is fully inelastic is assumed by \textit{Model 3}, in which the viscoplastic model presented in subsection \ref{subsec:viscoplastic} is employed. Finally, a combination of both hypotheses is assumed by \textit{Model 2}, where the viscoelastic element (Voigt element) is combined with the viscoplastic element. It should be stressed that the terminologies ``fully-elastic'', ``hybrid'' and ``fully-inelastic'', employed in Fig. \ref{fig:models_1_2_3} to designate the models, refer exclusively to the underlying assumption of each model regarding the transient creep stage. For instance, \textit{Model 1} is not fully elastic as a whole, but the transient creep stage is regarded as such.

\begin{figure}[!ht]
	\centering
	\includegraphics[scale=0.70]{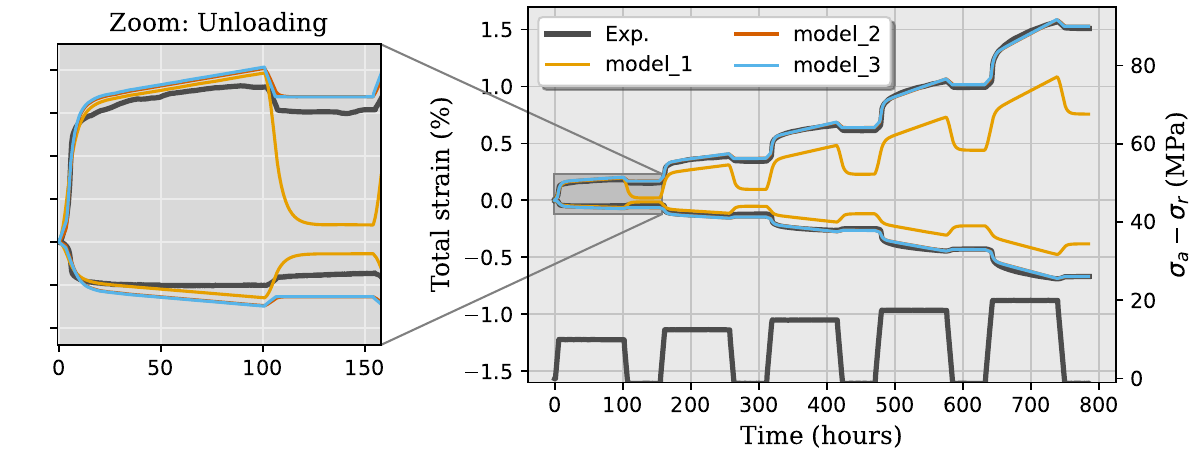}
	\caption{Total strain curves obtained with three different models.}
    \label{fig:BBG_Pr5_time_strain}
\end{figure}

\begin{figure}[!b]
	\centering
	\includegraphics[scale=0.70]{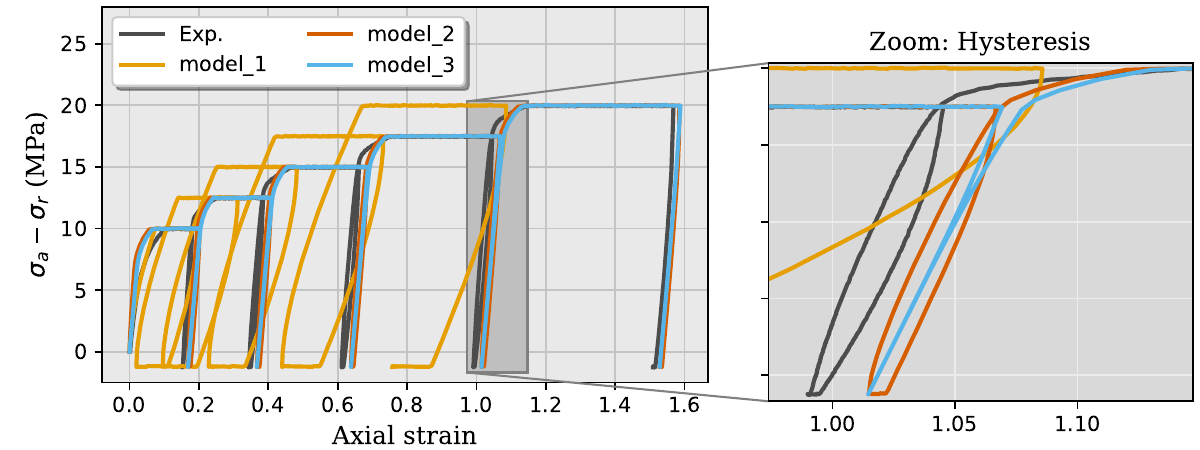}
	\caption{Stress-strain curves showing the unloading/loading step (reverse creep).}
    \label{fig:BBG_Pr5_stress_strain}
\end{figure}

The models illustrated in Fig. \ref{fig:models_1_2_3} are now used to fit laboratory experimental data. The experimental setup and salt rock details are presented in Section \ref{sec:lab_experiment}. It can be verified from Fig. \ref{fig:BBG_Pr5_time_strain} that \textit{Model 1} can be calibrated to properly describe the transient creep during the first step load. However, the assumption that transient creep strains are fully recoverable implies that most of the deformation should be reversed when the load is removed. This is precisely what we observe in the zoomed-in graph on the left of Fig. \ref{fig:BBG_Pr5_time_strain}, where the unloading takes place at around 100 hours. As expected, the total strain obtained with \textit{Model 1} is almost fully recovered after unloading, but this is not observed in the experimental results. In fact, only a small strain decrease is observed when the load is removed. Furthermore, the stress-strain graph in Fig. \ref{fig:BBG_Pr5_stress_strain} shows that \textit{Model 1} provides a completely wrong behavior. This shows that transient creep is not fully elastic, and therefore cannot be described by a simple Kelvin-Voigt element.

It can be verified in Fig. \ref{fig:BBG_Pr5_stress_strain} that Models 2 and 3 provide almost identical results, and both of them are able to describe the experimental data. In spite of that, the zoomed-in graph in Fig. \ref{fig:BBG_Pr5_stress_strain} clearly shows that \textit{Model 3} is unable to capture the reverse creep (hysteretic effect) observed during the unloading/reloading step. By contrast, the Kelvin-Voigt element of \textit{Model 2} can be tuned to properly describe reverse creep. 

\section{Calibration procedure}
In this section, we investigate the calibration procedure described in Section \ref{sec:calibration}. A set of synthetic experiments is designed to test the proposed strategy. The idea is to simulate a situation in which the experiments are provided one at a time, and the number of experiments available is not sufficient for a proper model calibration. In this context of high uncertainty, the goal is to perform a model calibration as good as possible with the experiments available and also to quantify the associated uncertainties.

\subsection{Synthetic experiments}
\label{subsec:synthetic_exps}
To overcome the lack of available laboratory experiments, in this work we consider 6 synthetic experiments in which the correct model parameters are known, thus providing the possibility of quantifying the calibration accuracy. The idea is to simulate a situation, where the experiments are made available one at a time, and the calibration takes place with the available data. This means that the calibration is first performed on Exp. \#1. Then, Exp. \#2 is made available and calibration is performed considering experiments \#1 and \#2. This is repeated until all six synthetic experiments are available.

\begin{figure}[!ht]
	\centering
	\includegraphics[scale=0.7]{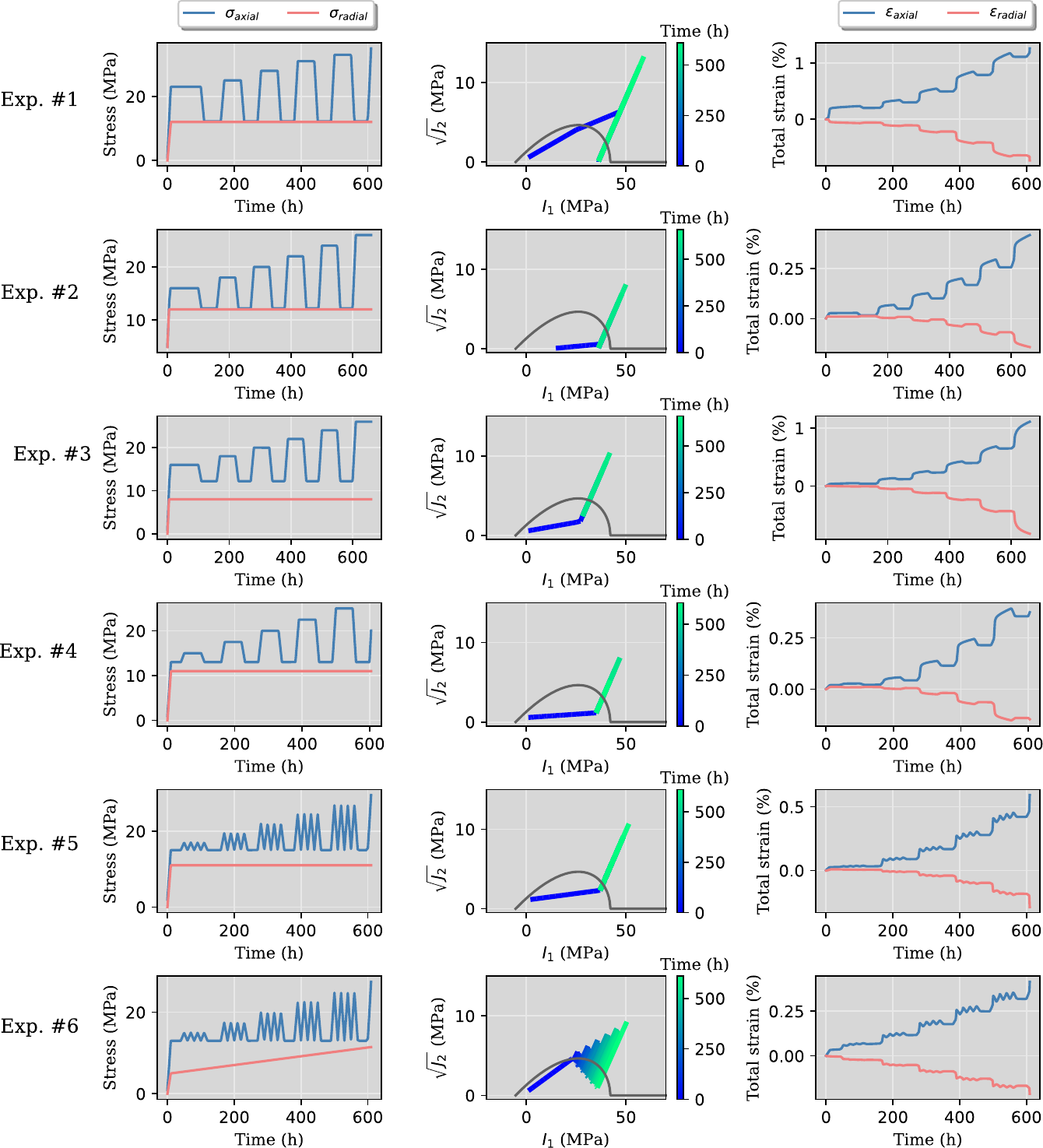}
	\caption{Synthetic experiments. Graphs on the left show the axial and radial stresses for each experiment. The middle column shows the corresponding stress path in the $I_1-\sqrt{J_2}$ plane, and the initial yield surface is also indicated according to Eq. \eqref{eq:F_vp} for $\alpha=\alpha_0$. The right column presents the resulting deformations of the synthetic sample.}
    \label{fig:paper_exps_0}
\end{figure}

The synthetic experiments are shown in Fig. \ref{fig:paper_exps_0}. The stress conditions were chosen as shown in the left column of the figure. The graphs in the middle column show the stress path of each experiment with the colors representing time. In these same graphs, the initial yield surface of the viscoplastic model is also plotted for reference, according to Eq. \eqref{eq:F_vp}. Finally, the corresponding strain responses are shown in the right column of Fig. \ref{fig:paper_exps_0}. In all stress conditions shown in Fig. \ref{fig:paper_exps_0}, the axial load is larger than the confining pressure, which means that the same Lode angle ($\theta=60^o$) was considered in all experiments. 

All experiments shown in Fig. \ref{fig:paper_exps_0} use the same set of material properties. The dislocation creep (Group 1) and viscoelastic (Group 2) parameters are shown in Tables \ref{tab:params_group_1} and \ref{tab:params_group_2}, respectively. For the viscoplastic element, the material parameters that are fixed (i.e., not obtained by calibration) are shown in Table \ref{tab:params_group_3}. Finally, Table \ref{tab:params_synthetic} presents the material parameters meant to be obtained through the calibration process. Even though these values are known, they are regarded as unknown so that we can test the optimization procedure.

\begin{table}[!ht]
    \centering
    \caption{Material properties for the synthetic experiments.}
    \begin{tabular}{ccc}
        \textbf{Parameter} & \textbf{Value}         & \textbf{Unit} \\ \hline
        $\mu_1$            & $6.898\times 10^{-12}$ & $s^{-1}$      \\
        $N_1$              & 3                      & $-$           \\
        $a_1$              & $1.8\times 10^{-5}$    & MPa$^{2-n}$   \\
        $\eta$             & $0.82$                 & $-$           \\
        $\alpha_0$         & $0.002$                & $-$          
    \end{tabular}
    \label{tab:params_synthetic}
\end{table}

\subsection{Global sensitivity analysis (GSA)}
Before proceeding with the model calibration, we investigate how the material parameters presented in Table \ref{tab:params_synthetic} affect the model results. For this purpose, we perform global sensitivity analysis (GSA), where all parameters are varied at the same time so that we can identify not just how each parameter affects the results individually, but also the interactions between them \cite{pianosi2016sensitivity}.

The strategy consists of generating a large population ($\sim$40 thousand) of material parameter sets. We build random uniform distributions for each material parameter within specified ranges, as shown in Table \ref{tab:prop_ranges}. For each set of parameters, we compute a loss function by comparing the resulting strains with a base solution. The loss function is computed as in Eq. \eqref{eq:loss_F}, with the set of parameters summarized in Table \ref{tab:params_synthetic} regarded as the base solution. In this manner, we can analyze how the changes in the material parameters affect the values of the loss function. 

\begin{table}[!ht]
    \centering
    \caption{Ranges of variation of each material property for the uniform distributions.}
    \begin{tabular}{r|ccccc}
        \textbf{Parameter} & $\log{\mu_1}$ & $N_1$ & $\log{a_1}$ & $\eta$ & $\log{\alpha_0}$ \\ \hline
        \textbf{Max.}      & -10           & 4     & -4          & 1.5    & -1.9             \\
        \textbf{Min.}      & -12           & 2     & -6          & 0.6    & -2.8  
    \end{tabular}
    \label{tab:prop_ranges}
\end{table}

\begin{figure}[!t]
	\centering
	\includegraphics[scale=0.7]{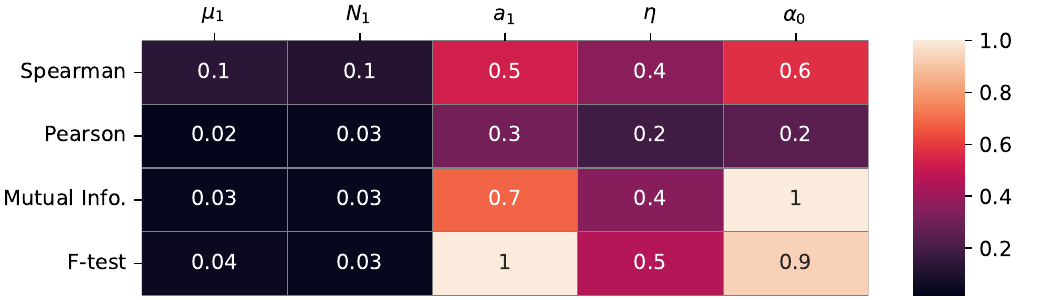}
	\caption{Table of correlation of each material parameter with the loss function. Darker and lighter colors refer to weaker and stronger correlations, respectively.}
    \label{fig:paper_corr_table}
\end{figure}

There are many ways GSA can be performed. One of them is computing statistical correlations between the material parameters and the loss function\footnote{We actually use the logarithm of the loss function as it spans many different orders of magnitude.}. The Pearson and Spearman correlations are commonly applied to identify linear and non-linear relationships, respectively, between variables \cite{rovetta2020raiders}. Additionally, univariate F-test is also able to capture linear relations, whereas mutual information can capture any type of dependency. It is important to stress that both Pearson's correlation and F-test assume normal distributions, which is not the case here. Nevertheless, the four methodologies are employed to identify relationships between the material parameters and the loss function. According to these methodologies, Fig. \ref{fig:paper_corr_table} reveals a strong relationship between the loss function and parameters $a_1$, $\eta$ and $\alpha_0$, the first two related to the hardening rule, and the last one defining the initial yield surface. All four methodologies point to a weak dependency on the rate-dependent parameters $\mu_1$ and $N_1$. These results are also confirmed by analyzing feature importance through a machine learning model and column permutation, as presented in \ref{app:feature_importance}.

\begin{figure}[!b]
	\centering
	\includegraphics[scale=0.7]{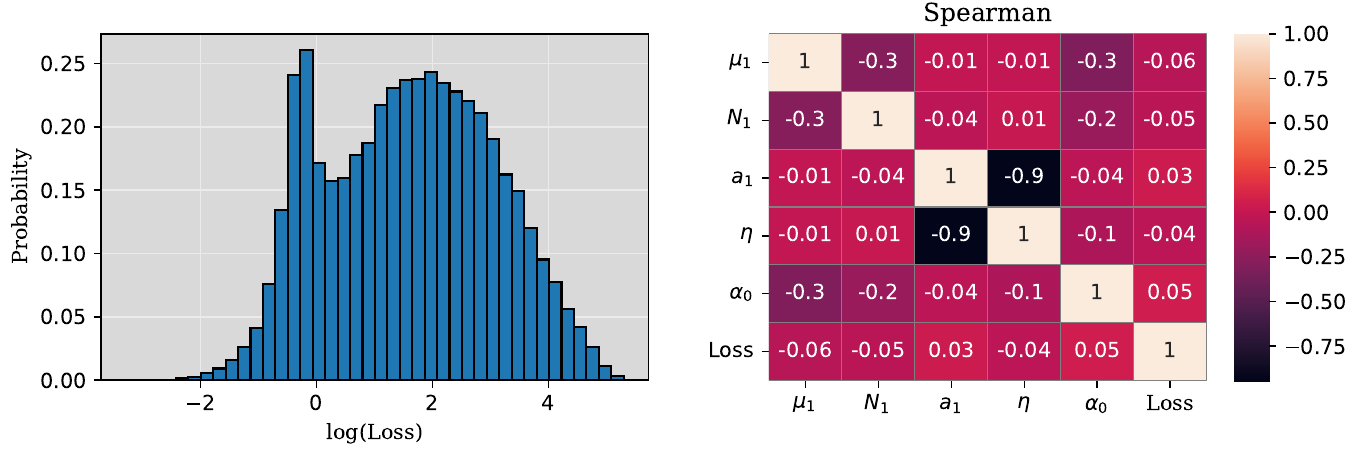}
	\caption{(Left) Loss function distribution. (Right) Spearman correlations for loss function values lesser than 0.1.}
    \label{fig:paper_loss_histogram}
\end{figure}

The dataset is created by choosing random values within the ranges specified in Table \ref{tab:prop_ranges}. As a consequence, the correlation between the material parameters is strictly zero. When the simulations are performed on the dataset, the resulting probability distribution of loss function values is shown in the left graph of Fig. \ref{fig:paper_loss_histogram}. We can see that there is a high probability of loss function values between 0.1 and 1.0, which suggests the existence of local minima. To further investigate this issue, we filter the loss function values larger than 0.1 out of our dataset and compute Spearman's correlations between all material parameters. In this manner, we only consider parameter sets that produce reasonably good results (loss function values smaller than 0.1). These correlations are shown in the right graph of Fig. \ref{fig:paper_loss_histogram}, where a strong relationship between $a_1$ and $\eta$ is identified. Weaker dependencies are also revealed between $\mu_1$, $N_1$ and $\alpha_0$. This shows that certain combinations between these variables can produce relatively good results.

\begin{figure}[!ht]
	\centering
	\includegraphics[scale=0.7]{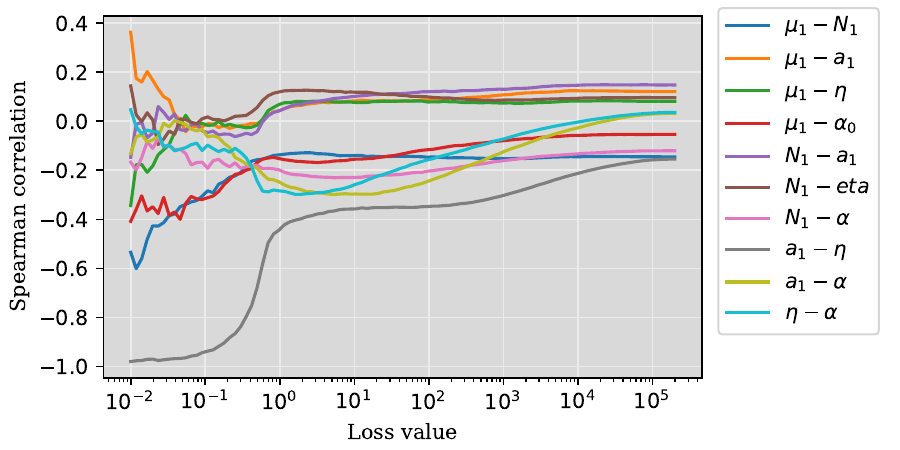}
	\caption{Spearman's correlation with filter out loss values.}
    \label{fig:paper_corr_corr_2}
\end{figure}

A better way to visualize this effect is by computing Spearman's coefficient as we continuously filter out the loss function values from $10^5$ down to $10^{-2}$. This is shown in Fig. \ref{fig:paper_corr_corr_2}, which clearly shows a significant increment of dependency between $a_1$ and $\eta$ for loss function values smaller than 1.0. To a lesser degree, similar behavior can be verified for the pairs $\mu_1-N_1$ and $\mu_1-\alpha_0$, as already revealed by the right graph of Fig. \ref{fig:paper_loss_histogram}. It is important to emphasize that all combinations of material parameters were obtained by random sampling, thus ensuring zero correlation between them. What the results of Figures \ref{fig:paper_loss_histogram} and \ref{fig:paper_corr_corr_2} reveal is that all sets of parameters that produce a relatively good solution present a certain (still unknown) correlation between $a_1$ and $\eta$. The sets of material parameters that do not present this correlation do not produce good results compared to the base case.

\begin{figure}[!ht]
	\centering
	\includegraphics[scale=0.70]{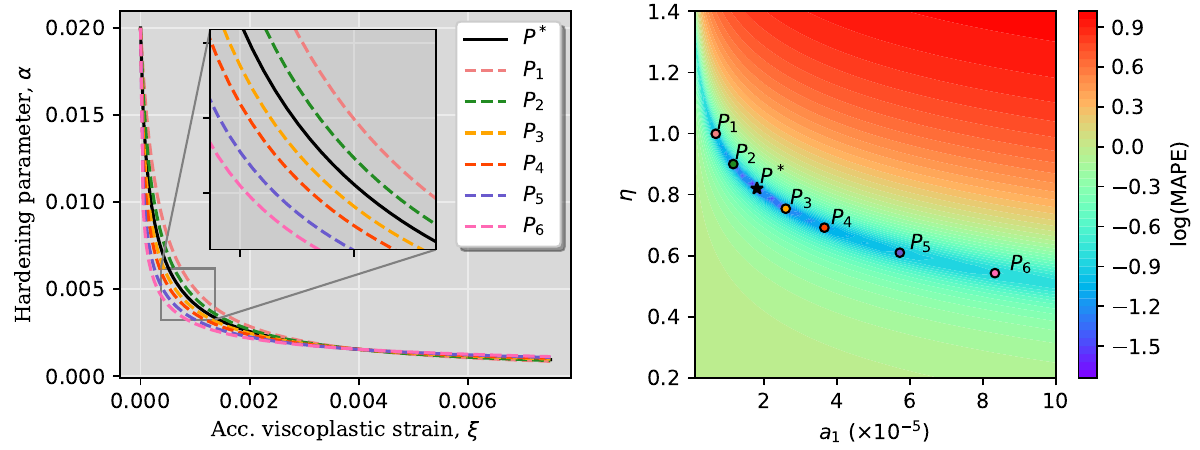}
	\caption{The left graph presents the behavior of $\alpha$ as a function of $\xi$ for different values of $\eta$ and $a_1$. The right graph shows the MAPE for each pair $\eta-a_1$ compared to the true value $P^*$.}
    \label{fig:paper_eta_a1_loss}
\end{figure}

To understand the origin of the correlation between $a_1$ and $\eta$ we use the hardening rule expressed by Eq. \eqref{eq:hardening}. In this equation, we consider $\alpha_0=0.02$, $\eta=0.82$ and $a_1=1.8\times 10^{-5}$. Then we vary the accumulated viscoplastic strain ($\xi$) from 0 to 0.0075, which are typical values. The obtained values for the hardening parameter are shown by the black curve on the left graph of Fig. \ref{fig:paper_eta_a1_loss}. This curve is regarded as a base case. Then we can choose different values of $a_1$ and $\eta$ and check how it affects the hardening parameter $\alpha$ with respect to $\xi$. For each pair $a_1-\eta$ we compute the MAPE and plot the color surface shown in the right graph of Fig. \ref{fig:paper_eta_a1_loss}. The true $a_1-\eta$ pair is indicated by $P^*$ in both graphs of Fig. \ref{fig:paper_eta_a1_loss}. We can see a region in which the values of $a_1$ and $\eta$ ($P_1$ to $P_6$) produce hardening curves very similar to the original $P^*$. In other words, whichever pair of values we take in this region, a good result may be produced by the constitutive model. Moreover, the existence of this region explains the high correlation between $a_1$ and $\eta$ in the right graph of Fig. \ref{fig:paper_loss_histogram} and Fig. \ref{fig:paper_corr_corr_2}.

The results obtained from the global sensitivity analysis, particularly the ones in Fig. \ref{fig:paper_eta_a1_loss}, give an idea about the behavior of the loss functions for the optimization procedure. These results show the presence of many local minima in which the optimization algorithm might be trapped, which allows the possibility of obtaining sub-optimal solutions.

\subsection{Single calibrations}
The goal of this subsection is to test the Particle Swarm Optimization (PSO) algorithm for finding the true material parameters of the synthetic experiments presented in Subsection \ref{subsec:synthetic_exps} (Fig. \ref{fig:paper_exps_0}). The idea is to run individual optimizations for each synthetic experiment and check whether the calibration process can indeed find the global solution, which is shown in Table \ref{tab:params_synthetic}. The hyperparameters used in the PSO algorithm (Eq. \ref{eq:pso_params}) are $\omega = 0.8$, $c_1 = 0.8$ and $c_2=0.4$, the algorithm is set to perform 100 iterations, and the population size is 1280. To maximize coverage of the search domain, Latin Hypercube Sampling (LHS) is adopted and the ranges of each material property are those presented in Table \ref{tab:prop_ranges}.

\begin{figure}[!ht]
	\centering
	\includegraphics[scale=0.7]{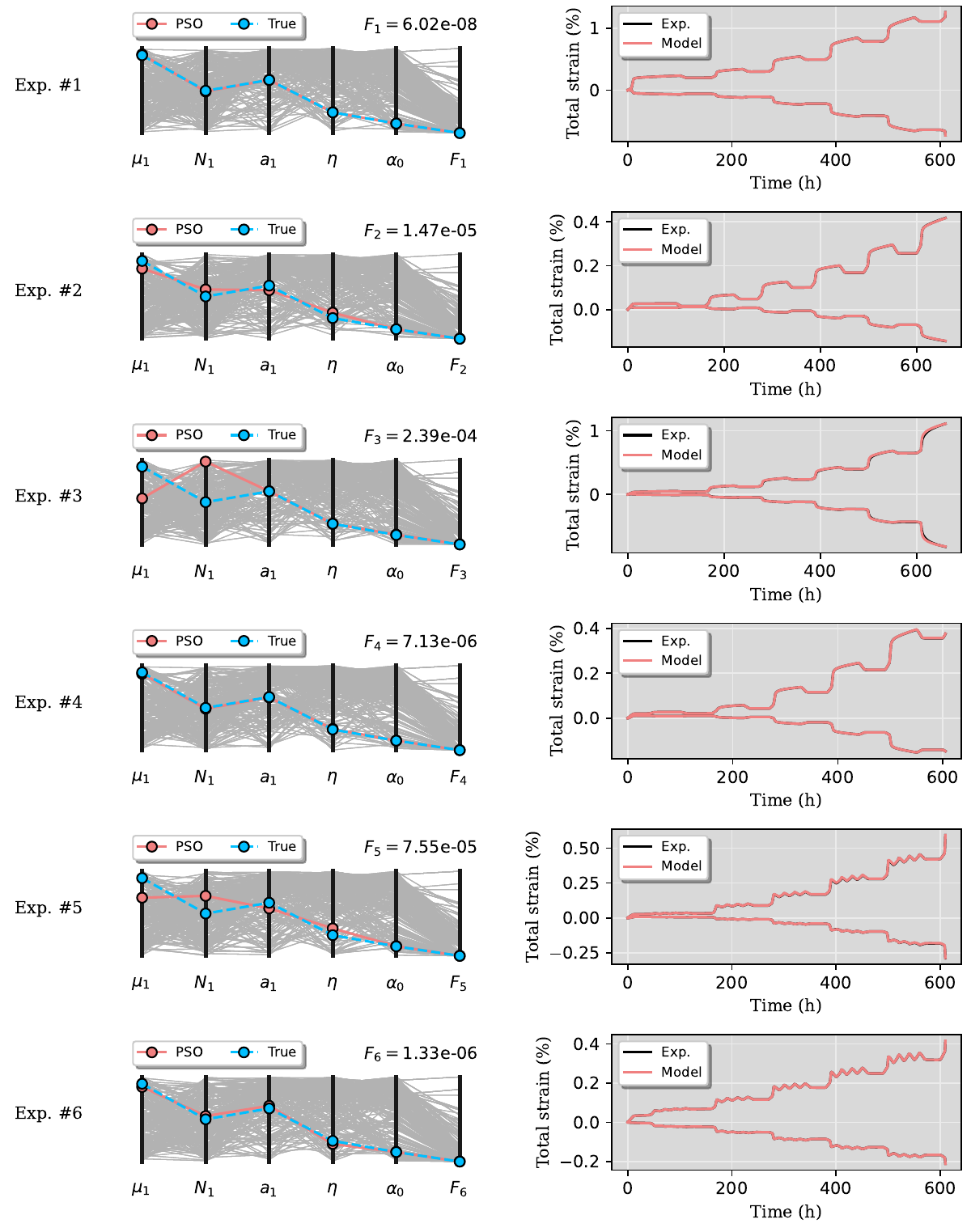}
	\caption{Synthetic experiments. (Left) Cobweb representation of the material parameters. (Right) Comparison between the true synthetic experiments and the calibrated model.}
    \label{fig:paper_single_0}
\end{figure}

The calibration outcomes obtained for each synthetic experiment are shown in Fig. \ref{fig:paper_single_0}. The left column of this figure shows a cobweb representation of the material properties. The gray lines represent the initial population, the red line denotes the set of properties obtained from the optimization process, and the true values are plotted in blue for comparison. The rightmost vertical bar in the cobwebs represents the loss function $F_i$ (MAPE), according to Eq. \ref{eq:loss_F}, and its corresponding values for the best solution are also indicated. The right column shows the total strains obtained with the calibrated properties for each experiment, and they all agree very well with the (synthetic) experimental results. For experiments \#3 and \#5 the values of $a_1$ and $N_1$ obtained from optimization differ from the true values. However, as revealed by the GSA, these two parameters have a secondary importance compared to the others, so these disparities do not have much impact on the behavior of the constitutive model. The values obtained from the calibrations of each experiment are summarized in Table \ref{tab:params_single}, along with the corresponding errors.

\begin{table}[!ht]
\centering
\caption{Material parameters obtained from the calibration process.}
\begin{tabular}{c|cccccc}
          & $\log{\mu_1}$ & $N_1$  & $\log{a_1}$ & $\eta$ & $\log{\alpha_0}$ & Loss ($F_i$)        \\ \hline
Reference & -11.1613      & 3.0000 & -4.7447     & 0.8200 & -2.6990          & 0.0                 \\
Exp. \#1  & -11.1414      & 2.9815 & -4.7450     & 0.8201 & -2.6991          & $6.02\times10^{-8}$ \\
Exp. \#2  & -11.3425      & 3.1648 & -4.8521     & 0.8772 & -2.6993          & $1.47\times10^{-5}$ \\
Exp. \#3  & -11.9118      & 3.9593 & -4.7436     & 0.8202 & -2.6951          & $2.39\times10^{-4}$ \\
Exp. \#4  & -11.1942      & 2.9800 & -4.7543     & 0.8256 & -2.6987          & $7.13\times10^{-6}$ \\
Exp. \#5  & -11.6245      & 3.4175 & -4.8789     & 0.8925 & -2.6972          & $7.54\times10^{-5}$ \\
Exp. \#6  & -11.2350      & 3.0760 & -4.6827     & 0.7875 & -2.6978          & $1.33\times10^{-6}$
\end{tabular}
\label{tab:params_single}
\end{table}

\subsection{Full calibrations}
In this test case, we aim to test the proposed calibration strategy in a more realistic scenario. In laboratory conditions, even if the rock samples are taken from the same region with a similar appearance, it is not possible to guarantee they will present exactly the same mechanical properties. Additionally, in general, the experiments are performed sequentially, meaning that they are not all available simultaneously. Considering that each creep experiment takes at least 20 to 30 days, approximately, it would be convenient to have partially calibrated models with the experiments at hand and include new experiments to the calibration as they become available.

\begin{figure}[!ht]
	\centering
	\includegraphics[scale=0.65]{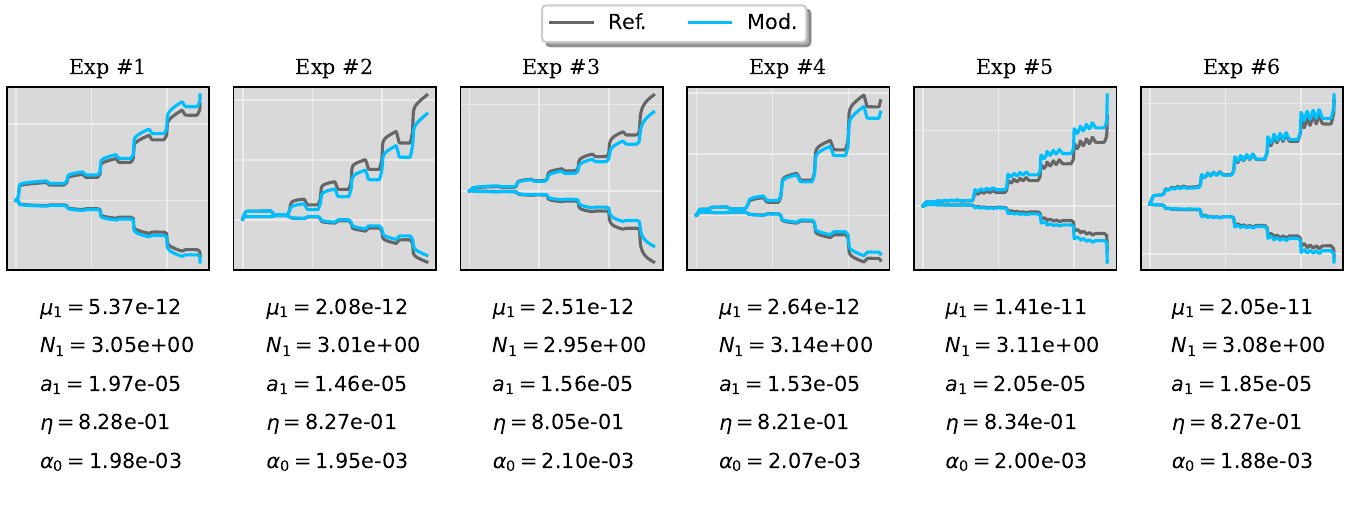}
	\caption{The black curves (Ref.) are obtained with the reference properties of Table \ref{tab:params_synthetic}, whereas the blue curves (Mod.) are created by perturbing (modifying) the reference properties. The modified material parameters are indicated below each graph.}
    \label{fig:paper_w4_1}
\end{figure}

To reproduce such a scenario, we take the reference values presented in Table \ref{tab:params_synthetic} and randomly disturb the material properties of each rock sample. In this manner, all experiments are performed with rock samples with slightly different material properties. The levels of perturbations are $\pm5\%$ for $\log{\mu_1}$ and $N_1$, $\pm2\%$ for $\log{a_1}$ and $\eta$, and $\pm1\%$ for $\log{\alpha_0}$. The material properties for the rock samples of each experiment are presented in Fig. \ref{fig:paper_w4_1}. The graphs in this same figure show the total strains obtained with the reference properties (i.e. those of Table \ref{tab:params_synthetic}) and the modified properties. It can be verified by visual comparison that the strain curves do not differ much from the reference case, but those small property variations do cause noticeable differences. Since the material properties of each experiment are different from each other, it is impossible to perfectly fit all experiments. Nevertheless, the goal of the calibration is to find a single set of material properties that fits the experiments as good as possible. Additionally, the calibration is performed in steps, including one new experiment at a time, according to Algorithm \ref{alg:alg_1}.

\begin{figure}[!t]
	\centering
	\includegraphics[scale=0.7]{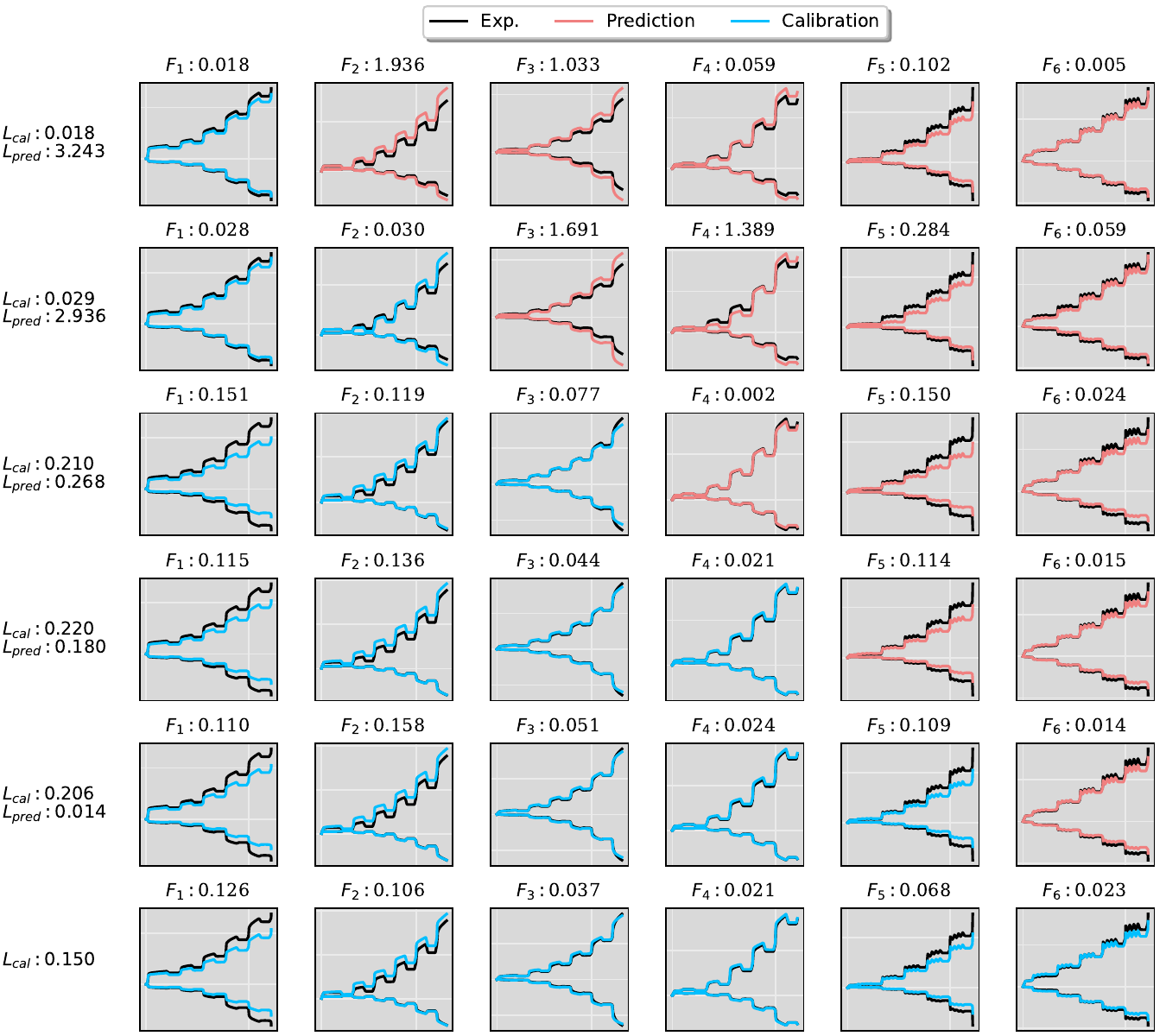}
	\caption{Calibration process including one experiment at a time.}
    \label{fig:paper_results_group_10}
\end{figure}

The result of the calibration procedure is presented in Fig. \ref{fig:paper_results_group_10}. The first line of this figure represents step 1, in which only experiment \#1 is available for calibration. Once the model is calibrated against experiment \#1, the model is used to predict the behaviors of the other experiments\footnote{Even though this would never be possible in real life, we use the remaining experiments for prediction in order to check predictability.}. Above each graph is indicated the $F_i$ error, as in Eq. \eqref{eq:loss_F}, for each experiment. The loss function values $L_i$, given by Eq. \eqref{eq:loss_L}, are presented on the left for the calibrated (blue) and predicted (red) experiments. In the first step, only experiment \#1 is available, so a single optimization (Eq. \ref{eq:w_F_i}) is performed using the initial population $\mathbb{P}$ to obtain $\mathbf{w}^F_1$. When experiment \#2 is available in the second step, another single optimization, with $\mathbb{P}$ as the initial population, is performed on it, which provides the set of parameters $\mathbf{w}^F_2$. Still on the second step, a combined optimization (Eq. \ref{eq:w_L_i}) is performed upon experiments \#1 and \#2 including $\mathbf{w}^F_1$ and $\mathbf{w}^F_2$ among the initial population $\mathbb{P}$. The combined optimization between experiments \#1 and \#2 provides $\mathbf{w}^L_2$, which should be included in the combined optimization of the subsequent steps as well. This whole process is summarized in Table \ref{tab:calibration_steps} for all calibration steps.

\begin{table}[!ht]
\centering
\caption{Population sets $\mathbb{P}$, $\mathbb{F}$ and $\mathbb{L}$ during the calibration steps.}
\begin{tabular}{c|cccc}
\textbf{Step} & \textbf{Initial population} & \textbf{$\mathbb{F}$}                                                                    & \textbf{$\mathbb{L}$}                                                  & \textbf{Output}                   \\ \hline
\#1           & $\mathbb{P}$                & $\emptyset$                                                                    & $\emptyset$                                                  & $\mathbf{w}^F_1$                   \\
\#2           & $\mathbb{P}$                & $\mathbf{w}^F_1$                                                                         & $\emptyset$                                                  & $\mathbf{w}^F_2$, $\mathbf{w}^L_2$ \\
\#3           & $\mathbb{P}$                & $\mathbf{w}^F_1$, $\mathbf{w}^F_2$                                                       & $\mathbf{w}^L_2$                                                       & $\mathbf{w}^F_3$, $\mathbf{w}^L_3$ \\
\#4           & $\mathbb{P}$                & $\mathbf{w}^F_1$, $\mathbf{w}^F_2$, $\mathbf{w}^F_3$                                     & $\mathbf{w}^L_2$, $\mathbf{w}^L_3$                                     & $\mathbf{w}^F_4$, $\mathbf{w}^L_4$ \\
\#5           & $\mathbb{P}$                & $\mathbf{w}^F_1$, $\mathbf{w}^F_2$, $\mathbf{w}^F_3$, $\mathbf{w}^F_4$                   & $\mathbf{w}^L_2$, $\mathbf{w}^L_3$, $\mathbf{w}^L_4$                   & $\mathbf{w}^F_5$, $\mathbf{w}^L_5$ \\
\#6           & $\mathbb{P}$                & $\mathbf{w}^F_1$, $\mathbf{w}^F_2$, $\mathbf{w}^F_3$, $\mathbf{w}^F_4$, $\mathbf{w}^F_5$ & $\mathbf{w}^L_2$, $\mathbf{w}^L_3$, $\mathbf{w}^L_4$, $\mathbf{w}^L_5$ & $\mathbf{w}^F_6$, $\mathbf{w}^L_6$
\end{tabular}
\label{tab:calibration_steps}
\end{table}

\begin{figure}[!ht]
	\centering
	\includegraphics[scale=0.65]{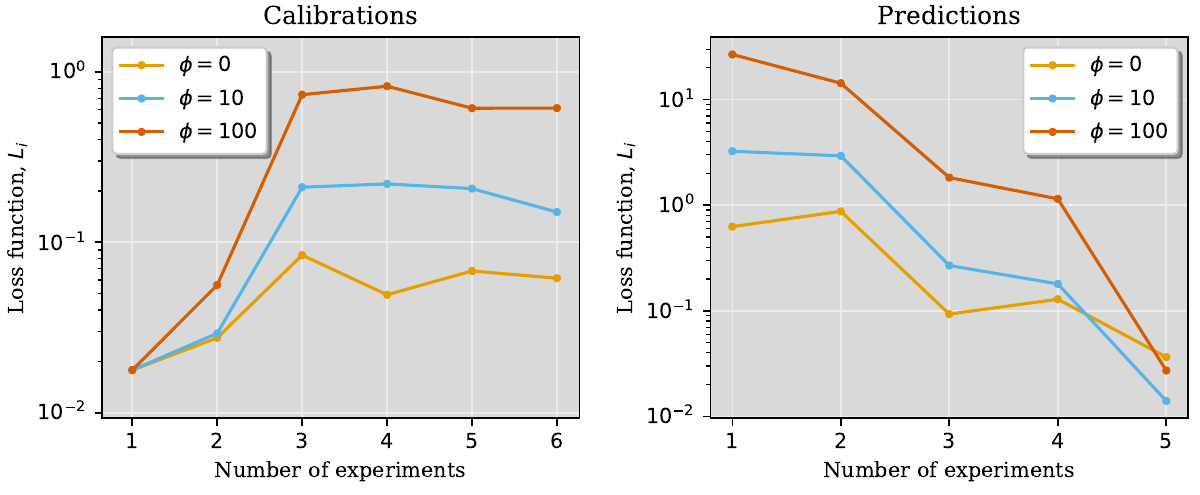}
	\caption{Loss function values for calibration and predictions.}
    \label{fig:paper_cal_pred}
\end{figure}

It can be verified from Fig. \ref{fig:paper_results_group_10} that the combined loss function for the calibrated experiments, $L_{cal}$, substantially increases as more experiments are included in the calibration procedure. At the same time, the $L_{pred}$ is observed to decrease, which suggests that the predictability of the calibrated constitutive model improves by adding new experiments. Figure \ref{fig:paper_cal_pred} plots these two quantities by the number of experiments available for calibration. On the left graph of this figure, we see that the loss function indeed increases in the beginning but it remains more or less constant as more experiments are added to the calibration. The reason for this is, as mentioned before, it is impossible to simultaneously fit all experiments since they have different material properties, so after a certain point the error is expected to stabilize at a constant level. On the left graph, however, we see that the more experiments we consider for calibration, the better the predictions of the constitutive model.

\begin{figure}[!ht]
	\centering
	\includegraphics[scale=0.65]{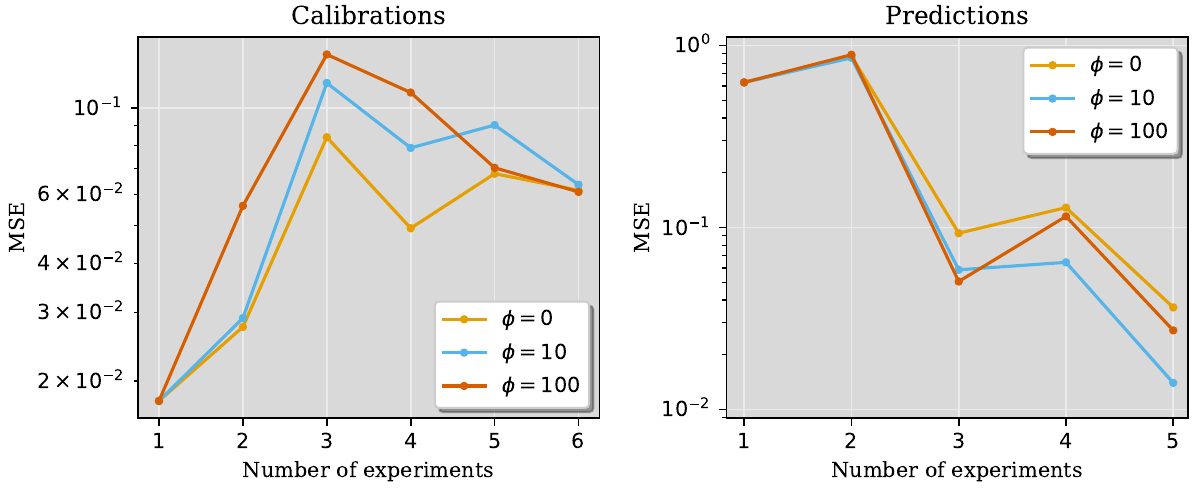}
	\caption{Mean squared errors (MSE) for calibration and predictions.}
    \label{fig:paper_cal_pred_2}
\end{figure}

Equation \eqref{eq:loss_L} includes the parameter $\phi$ as a regularization in the loss function for the combined optimization. The results presented in Fig. \ref{fig:paper_results_group_10} were obtained with a regularization factor of $\phi=10$. To investigate how the regularization term affects the calibration process, the same test case is run considering $\phi=0$, $\phi=10$ and $\phi=100$. The loss function values $L_i$ are shown in Fig. \ref{fig:paper_cal_pred} as a function of the number of experiments considered for calibration. The left graph shows the results of the calibration experiments, where the loss function increases until the third experiment is added and stabilizes at approximately constant values. As expected, the larger $\phi$ the larger $L_i$ since the regularization term becomes larger. The graph on the right shows the loss function values for the experiments left for prediction. It can be seen that the loss function values decrease as more experiments are considered for calibration. In other words, it shows that the constitutive model can make better predictions when more experiments are considered during calibration. Additionally, it can be verified that for 5 experiments considered during calibration, the prediction of the 6th experiment is better when $\phi=10$. Since it might be difficult to compare the different curves in Fig. \ref{fig:paper_cal_pred} because of the regularization term, in Fig. \ref{fig:paper_cal_pred_2} only the MSE is shown. In this case, the comparisons are unambiguously and we can see from the right graph that indeed better predictions are obtained with $\phi=10$. 

\begin{figure}[!ht]
	\centering
	\includegraphics[scale=0.70]{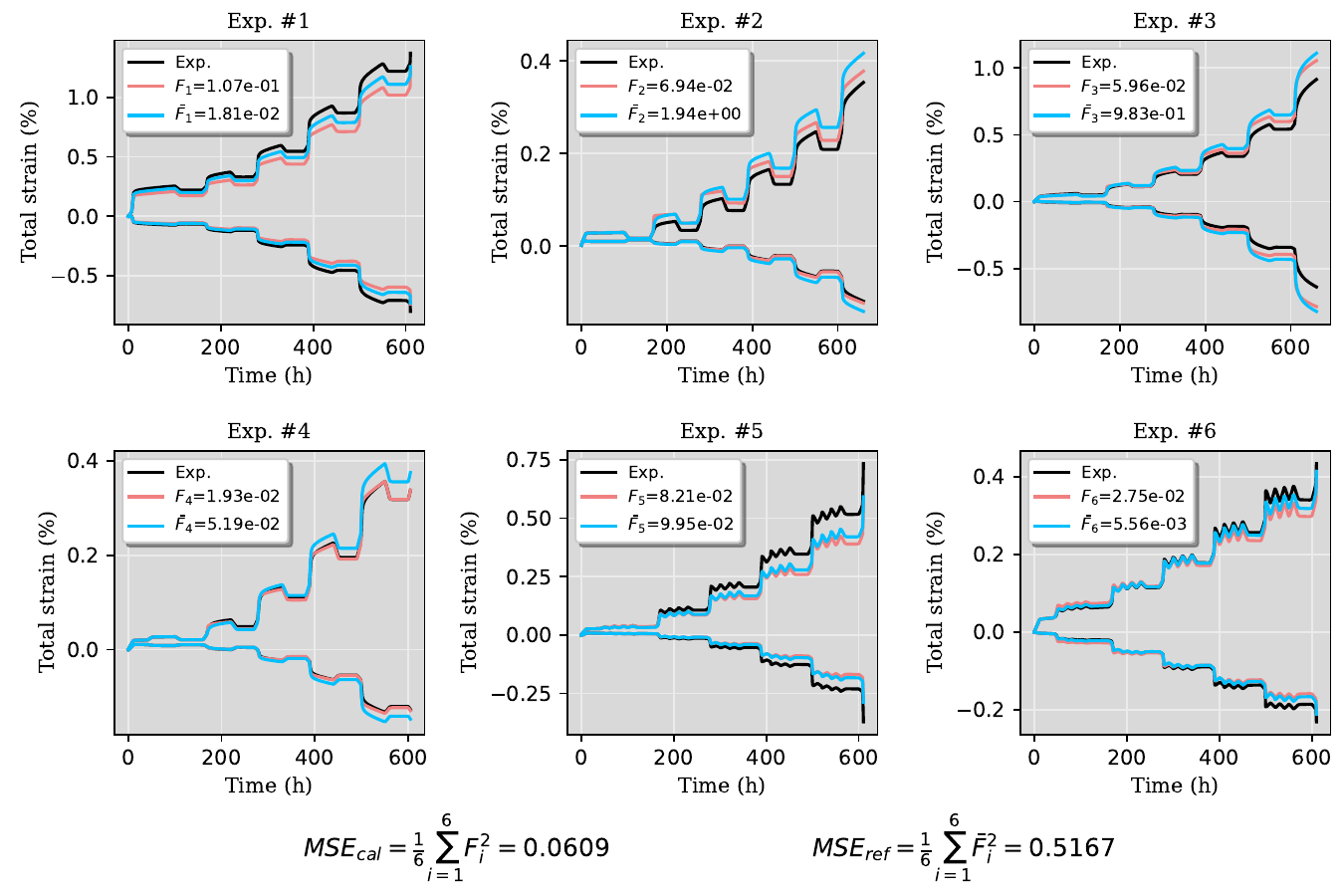}
	\caption{Constitutive model results obtained with the material properties obtained from calibration and the reference values from Table \ref{tab:params_synthetic}. The overbar refers to the quantities related to the reference parameters.}
    \label{fig:paper_mean_0}
\end{figure}

As explained before, the sets of properties chosen for each experiment were obtained by randomly disturbing the reference properties of Table \ref{tab:params_synthetic}. Therefore, it would be reasonable to expect that the reference properties themselves would also provide good performance in fitting all six experiments as well. In Fig. \ref{fig:paper_mean_0} we compare the total strains for all experiments considering the calibrated (red curves) and the reference (blue curves) set of material properties. The curves in black represent the true values of the synthetic experiments. It can be verified that experiments \#2, \#3 and \#4 are better fitted by the calibrated parameters, whereas the reference properties perform slightly better for experiments \#1, \#5 and \#6. However, the mean squared errors (MSE) shown at the bottom of this figure reveal that the calibrated material parameters present an overall better fit.

The calibrated results presented in Fig. \ref{fig:paper_mean_0} are obtained with a regularization parameter of $\phi=100$. Different regularization parameters, however, may lead to different calibrated material properties. The obtained values for $\phi$ equals to 0, 10 and 100 are summarized in Table \ref{tab:phis}, along with the reference properties for comparison. As shown in this table, the MSE values of the calibrated properties are always smaller than the reference properties. Interestingly, the material properties obtained through the calibration process are significantly different from each other and the reference values. Apart from $\alpha_0$, the other material properties even lie outside the limits of the perturbed properties, as it can be verified in Fig. \ref{fig:paper_w4_1}. This can be attributed to the non-linearity of the viscoplastic model.

\begin{table}[!ht]
\centering
\caption{Calibrated material properties obtained with different values of $\phi$, and the reference properties from Table \ref{tab:params_synthetic} for comparison. The last line shows the mean squared error (MSE) for each set of properties.}
\begin{tabular}{c|cccc}
           & $\phi=0$             & $\phi=10$            & $\phi=100$           & Reference            \\ \hline
$\mu_1$    & $4.51\times10^{-12}$ & $1.83\times10^{-12}$ & $9.77\times10^{-12}$ & $6.89\times10^{-12}$ \\
$N_1$      & 2.98                 & 3.99                 & 3.47                 & 3.00                 \\
$a_1$      & $4.33\times10^{-5}$  & $3.12\times10^{-6}$  & $2.19\times10^{-5}$  & $1.80\times10^{-5}$  \\
$\eta$     & 0.60                 & 1.19                 & 0.74                 & 0.82                 \\
$\alpha_0$ & 0.00207              & 0.00205              & 0.00202              & 0.0020               \\
MSE        & 0.0615               & 0.0635               & 0.0609               & 0.5167              
\end{tabular}
\label{tab:phis}
\end{table}

\section{Conclusion}
\label{sec:conclusion}
Geomechanics of salt caverns for safe cyclic hydrogen storage requires a detailed investigation of the associated creep mechanisms. For this purpose, we assemble a constitutive model able to capture transient creep, reverse creep and steady-state dislocation creep. This is achieved by including elastic, viscoelastic, viscoplastic, and power-law contributions to the constitutive model. The viscoelastic contribution can capture the hysteretic effect that appears during unloading/reloading stages, also known as reverse creep. Moreover, it is shown that the viscoelastic element alone is not able to appropriately describe transient creep, for which a viscoplastic model is indispensable. With this constitutive model, the experimental results presented in this paper can be successfully described.

An important step in devising and deploying a constitutive model relates to model calibration. Due to the time scales involved in creep deformations, laboratory experiments in salt rocks are usually very time-consuming (approximately one month long or even more). Additionally, it takes a lot of experiments to properly calibrate a constitutive model. There is also the possibility that mechanical properties might vary from sample to sample, making the calibration process more challenging. In this context, the main contribution of this paper is to devise a multi-step calibration strategy that takes place on the fly, in the sense that the constitutive model is calibrated against experiments as they are made available.
Determination of the material parameters corresponding to the elastic, viscoelastic, and dislocation creep can be done independently because the presented procedure is general and flexible. Another significant contribution of the present study is on the calibration of the viscoplastic material parameters, which have to be all tuned simultaneously. Moreover, the choice of the material parameters to be calibrated is discussed, and a systematic analysis for the impact of these parameters on the constitutive model behavior is also presented. The calibration strategy consists of sequentially solving multi-objective optimization problems as new experiments are included. 

It is shown that the proposed strategy provides increasingly better results as more experiments are incorporated into the calibration process. In this manner, calibrated constitutive models are provided much more efficiently and reliably than the existing procedures in the literature. Additionally, it is shown that including a regularization term into the loss function is beneficial to promote equal quality fit of the constitutive model against all experimental data sets. The proposed methodology can be applied to any constitutive model, and it provides representative material parameters for salt rocks. As such, this work contributes to improving the accuracy of constructing a suitable constitutive model with reliable material parameter sets to study the mechanics of salt caverns under cyclic operations. Future studies will incorporate this characterization procedure in field-scale 3D cavern simulations.

\section*{CRediT authorship contribution statement}
 
\textbf{H.T.H.}: Conceptualization, Methodology, Software, Validation, Formal analysis, Visualization, Writing -- Original Draft.
\textbf{M.H.}: Conceptualization, Investigation, Writing -- Review \& Editing.
\textbf{K.B.}: Conceptualization, Writing -- Review \& Editing.
\textbf{A.L.}: Investigation.
\textbf{K.B.}: Conceptualization, Writing -- Review \& Editing.
\textbf{L.S}:  Conceptualization, Methodology, Writing -- Review \& Editing
\textbf{H.H.}: Conceptualization, Methodology, Writing -- Review \& Editing.

\section*{Declaration of competing interest}
The authors declare that they have no known competing financial interests or personal relationships that could have appeared to influence the work reported in this paper.

\section*{Data availability}
Digital datasets of the results and input data are available upon request.

\section*{Acknowledgments}
This research was partly supported by Shell Global Solutions International B.V within the project `SafeInCave'. The authors also acknowledge members of the ADMIRE and DARSim research groups at TU Delft for the fruitful discussions during the development of this work. Additionally, the authors acknowledge the fruitful discussions with the technical staff of Shell Global Solutions International B.V and for providing the experimental data set used in this work.

\appendix

\section{Feature importance for sensitivity analysis}
\label{app:feature_importance}
\begin{figure}[!ht]
	\centering
	\includegraphics[scale=0.7]{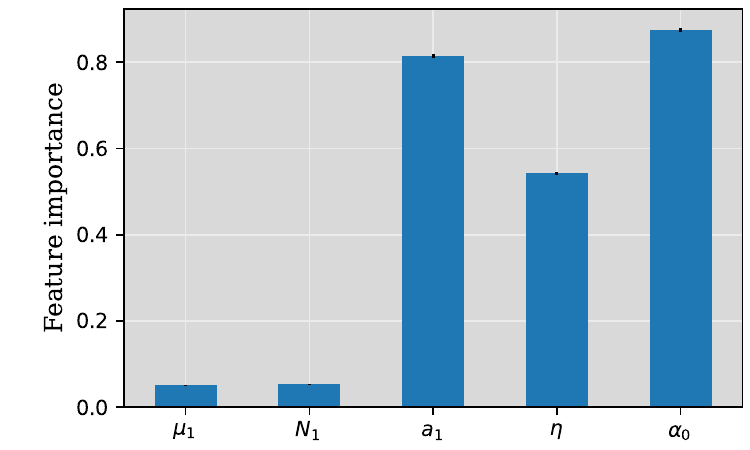}
	\caption{Feature importance according to the column permutation method.}
    \label{fig:paper_feature_importance}
 \end{figure}

Another interesting strategy to measure the influence of each material parameter on the model behavior is through feature importance, as usually performed in machine learning (ML) applications. Some ML models, such as linear regression and decision trees, naturally provide feature importance rankings. A more general approach can be achieved by column permutation. This strategy consists of training a ML model to predict the target variable (in our case, the loss function). After the training step, the model performance is evaluated according to a certain metric (e.g. MSE). This establishes a base metric for the ML model. Then, we shuffle the values of a certain feature (i.e., material parameter), run the model again without retraining, and check the new metric. Based on how much the new metric is reduced compared to the base metric we can infer the importance of that feature. A great reduction means the ML model heavily relies on that feature. Conversely, if the metric barely changes, it means the ML model does not find that feature very useful. This procedure is performed using the Random Forest algorithm and the results are shown in Fig. \ref{fig:paper_feature_importance}. We can see that parameters $\mu_1$ and $N_1$ are of less importance compared to $a_1$, $\eta$ and $\alpha_0$. This is in complete agreement with the correlations found in Figure \ref{fig:paper_corr_table}.

\bibliographystyle{unsrt}
\bibliography{main}

\end{document}